%% file: paper.tex
\documentclass[%
 reprint,
 superscriptaddress,preprintnumbers,
 amssymb,
 aps,
]{revtex4-2}

\usepackage[utf8]{inputenc}
\usepackage{hyperref}
\usepackage{amsmath,amssymb,mathtools}
\usepackage{dsfont}
\usepackage{graphicx}   
\usepackage[table]{xcolor}
\usepackage{colortbl}
\usepackage{url}
\usepackage[normalem]{ulem}
\usepackage{slashed}
\usepackage[capitalise, english]{cleveref}
\usepackage[noindentafter]{titlesec} 
\allowdisplaybreaks

\usepackage{multirow}
\usepackage{booktabs}
\usepackage[export]{adjustbox}

\usepackage{siunitx}
\usepackage{xspace}

\usepackage{float}

    \makeatletter
    \def\CT@@do@color{%
      \global\let\CT@do@color\relax
            \@tempdima\wd\z@
            \advance\@tempdima\@tempdimb
            \advance\@tempdima\@tempdimc
    \advance\@tempdimb\tabcolsep
    \advance\@tempdimc\tabcolsep
    \advance\@tempdima2\tabcolsep
            \kern-\@tempdimb
            \leaders\vrule
                    \hskip\@tempdima\@plus  1fill
            \kern-\@tempdimc
            \hskip-\wd\z@ \@plus -1fill }
    \makeatother

\usepackage{accents}
\DeclareMathSymbol{\widetildesym}{\mathord}{largesymbols}{"65}

\titleformat*{\subsubsection}{\normalfont \small \bfseries \boldmath}
\renewcommand{\paragraph}[1]{\vspace{.3em} \indent {\bfseries \boldmath #1 ---}\xspace }
\makeatletter
    \renewcommand{\p@subsection}{}
    \renewcommand{\p@subsubsection}{}
\makeatother

\definecolor{red}{rgb}{0.6,.0706,.1373}
\definecolor{blue}{rgb}{0,0.396,0.741}
\newcommand\myshade{80}
\colorlet{mylinkcolor}{violet}
\colorlet{mycitecolor}{violet}
\colorlet{myurlcolor}{violet}

\hypersetup{
  linkcolor  = mylinkcolor!\myshade!black,
  citecolor  = mycitecolor!\myshade!black,
  urlcolor   = myurlcolor!\myshade!black,
  colorlinks = true
}
\newcommand{\cO}{\mathcal{O}}

\newcommand{\diag}{\mathop{\mathrm{diag}}}

\newcommand{\eminus}{\vcenter{\hbox{\scalebox{0.6}[1]{$ - $}}}}	

\newcommand{\sscript}[1]{{\scriptscriptstyle \mathrm{#1}}}


\begin{document}


\title{
 \boldmath %
Minimal Flavor Protection for TeV-scale New Physics
}

\author{Admir Greljo}
\email{admir.greljo@unibas.ch}
\affiliation{Department of Physics, University of Basel, Klingelbergstrasse 82, CH-4056 Basel, Switzerland}

\author{Ajdin Palavri\' c}
\email{ajdin.palavric@ific.uv.es}
\affiliation{Department of Physics, University of Basel, Klingelbergstrasse 82, CH-4056 Basel, Switzerland}
\affiliation{Instituto de F\'isica Corpuscular (IFIC), Consejo Superior de Investigaciones
Cient\'ificas (CSIC) and Universitat de Val\`encia (UV), 46980 Valencia, Spain}

\author{Ben A. Stefanek}
\email{bstefan@ific.uv.es}
\affiliation{Instituto de F\'isica Corpuscular (IFIC), Consejo Superior de Investigaciones
Cient\'ificas (CSIC) and Universitat de Val\`encia (UV), 46980 Valencia, Spain}



\begin{abstract}

We determine how much TeV-scale new physics can deviate from flavor universality, $U(3)^5$, while respecting stringent bounds on flavor-changing neutral currents. The minimal continuous subgroup that \emph{must} be approximately preserved is found to be of the form $SU(2)_{q} \times U(1)_{X}$. With only a few symmetry-breaking spurions of $\mathcal{O}(10^{-2})$, all observed fermion hierarchies may be reproduced, offering a new perspective on the flavor puzzle. Remarkably, this framework provides structural flavor protection for generic TeV-scale new physics within the SMEFT, enlarging the space of collider-accessible scenarios beyond MFV and $U(2)^5$ and allowing for richer patterns of flavor violation.

\end{abstract}

\maketitle

\section{Introduction} 
\label{sec:intro}

Mass hierarchies are among the most profound structural questions in particle physics and call for an explanation beyond the Standard Model (BSM). A striking example is the peculiar hierarchical pattern of fermion masses and mixings, the \textit{SM flavor puzzle}~\cite{Feruglio:2015jfa, Altmannshofer:2024hmr}, which points to the existence of an underlying structure behind the observed Yukawa interactions of the Higgs field. Being dimensionless, the Yukawa couplings do not single out a characteristic scale at which this structure must emerge. In contrast, the electroweak hierarchy problem, arising from the sensitivity of the Higgs mass to ultraviolet (UV) dynamics, suggests that new physics (NP) should appear not far above the TeV scale, the present energy frontier of collider experiments. 

The hypothesis of TeV-scale, collider-accessible NP with sizable couplings to SM fields is well motivated and actively pursued in ongoing searches at the Large Hadron Collider (LHC), but it gives rise to a new puzzle, the \textit{NP flavor puzzle}~\cite{Altmannshofer:2024hmr, Isidori:2025iyu, Nir:2020jtr, Altmannshofer:2025rxc, Buras:2020xsm, Zupan:2019uoi, Isidori:2010kg}. Stringent indirect bounds from flavor-changing neutral currents (FCNC) and CP-violating observables imply that such NP cannot have a generic flavor structure, but must exhibit special protection mechanisms to evade current limits from low-energy precision experiments.

Without Yukawa interactions, the SM has a large global flavor symmetry, $U(3)^5$~\cite{Chivukula:1987py}, corresponding to independent unitary rotations of the three generations of each fermion gauge representation. The Yukawa couplings explicitly break this symmetry, leaving only baryon number and individual lepton flavors, $U(1)_B \times U(1)^3$, as exact symmetries. Owing to the hierarchical structure, one can identify approximate flavor symmetries of increasing quality: the largest breaking, driven by $y_t \sim \mathcal{O}(1)$, reduces the original symmetry down to $U(2)^2 \times U(1) \times U(3)^3$, followed by a smaller breaking of $\mathcal{O}(0.01)$, etc.

Mechanisms that protect TeV-scale NP from excessive flavor violation ensure that approximate flavor and CP symmetries of the SM are also respected in the NP sector. This predicts BSM selection rules that mirror those of the SM, keeping flavor and CP observables under control. The classic paradigm is \textit{Minimal Flavor Violation} (MFV)~\cite{DAmbrosio:2002vsn, Cirigliano:2005ck}, where the three SM Yukawa matrices serve as spurions that encode all sources of $U(3)^5$ and CP violation, including those in the BSM sector. Among the variations and alternatives~\cite{Feldmann:2008ja, Kagan:2009bn, Agashe:2005hk, Egana-Ugrinovic:2018znw}, the most extensively studied framework is the minimally-broken $U(2)^5$ flavor symmetry~\cite{Barbieri:2011ci, Barbieri:2012uh,Fuentes-Martin:2019mun}. It distinguishes the third family from the first two, reproduces the SM Yukawa structure with four spurions, and simultaneously provides flavor protection for NP. Within the Standard Model Effective Field Theory (SMEFT)~\cite{Grzadkowski:2010es, Isidori:2023pyp, Aebischer:2025qhh}, used as a proxy for generic short-distance NP, both flavor symmetries and their breakings can be implemented order by order in the spurion expansion~\cite{Greljo:2022cah, Faroughy:2020ina}. The resulting power counting provides structural suppression of dangerous FCNC.

The relationship between the two paradigms yields valuable insights. An $\mathcal{O}(1)$ breaking of $U(3)^5$ down to a smaller $U(2)^5$ symmetry allows for TeV-scale NP which exhibits richer flavor phenomenology, particularly in the $b$ and $\tau$ sectors~\cite{Fuentes-Martin:2019mun, Allwicher:2023shc}. In addition, the $U(2)^5$ framework provides a setting to (partially) address the SM flavor puzzle, since, unlike MFV, the Yukawas themselves originate from small spurions.

While MFV guarantees sufficient flavor protection, it is not strictly necessary and is thus overly restrictive: it excludes a broad class of potentially interesting flavor phenomena that could still be compatible with TeV-scale dynamics. Following the success of $U(2)^5$ as a first step in this direction, we ask \textit{how much} TeV-scale NP can \textit{further} break the full $U(3)^5$ flavor symmetry at $\mathcal{O}(1)$ without conflicting with stringent bounds from FCNC and CP-violating observables.  A pragmatic requirement is only that the most dangerous observables, such as $K$–$\bar K$, $D$–$\bar D$ mixing, $\mu \to e \gamma$, and electric dipole moments (EDMs) receive strong suppression, as these processes probe energies far above the TeV scale. 

We formulate this question by searching for a linearly realized, connected, compact subgroup $\mathcal{H}\subset U(3)^5$, with the smallest-dimensional Lie algebra, that must remain a good approximate symmetry at the TeV scale, while all other directions in flavor space are allowed to be maximally violated, i.e.~an $\mathcal{O}(1)$ breaking $U(3)^5\to\mathcal{H}$. The residual symmetry $\mathcal{H}$ is broken by an economical set of small spurions, whose sizes control and reproduce the observed fermion masses and mixings. We use the baryon-number-conserving dimension-six SMEFT as our proxy for generic short-distance NP. The symmetry--spurion structure sets the power counting of flavor violation, either through explicit spurion insertions or through rotations to the mass basis. We seek the minimal symmetry--spurion framework that protects generic TeV-scale NP from dangerous FCNC while reproducing the SM flavor parameters.

In summary, we introduce no more flavor suppression than necessary. This framework, which we name \emph{Minimal Flavor Protection} (MFP)~\footnote{The nomenclature MFP was previously used in the narrow context of warped extra dimensions~\cite{Santiago:2008vq}.} as a structural alternative to MFV and $U(2)^5$, has immediate implications for ongoing flavor experiments at bottom, charm, tau, muon and kaon facilities~\cite{LHCb:2018roe, Belle-II:2018jsg, Belle-II:2022cgf, LHCb:2021glh, BESIII:2020nme, MEGII:2018kmf, Bernstein:2019fyh, Hesketh:2022wgw, Moritsu:2022lem, NA62KLEVER:2022nea, Aebischer:2025mwl}, EDMs~\cite{ACME:2018yjb, Wu:2019jxj, n2EDM:2021yah} as well as high-$p_T$ flavor studies at the LHC~\cite{Greljo:2017vvb, Allwicher:2022gkm} and future colliders~\cite{deBlas:2025gyz, Greljo:2024ytg, Glioti:2025zpn, Allwicher:2025bub}. MFP generalizes MFV and $U(2)^5$, allowing for maximal violations of flavor universality and a richer pattern of flavor violation in a controlled manner, thereby providing a framework for systematic searches for TeV-scale flavor dynamics in the precision era.

Another motivation for reducing the starting flavor symmetry group is a tendency to transition from merely accommodating SM flavor hierarchies to explaining them within a dynamical framework. In this regard, even $U(2)^5$ is only a partial step: it leaves unexplained the hierarchical singular values of the bifundamental spurions $\Delta_{u,d,e}$ that generate masses for the first two generations, while MFV leaves the SM flavor hierarchies entirely unexplained. As we will show, in our framework all four spurions naturally appear at the same order, $\mathcal{O}(0.01)$, with no hierarchies between them nor within their components. This uniformity greatly simplifies the task of identifying a dynamical origin of these spurions, \emph{e.g.}, through the minimization of a scalar potential. This is known to be a non-trivial task for larger symmetry groups~\cite{Grinstein:2010ve, Alonso:2011yg, Nardi:2011st, Buras:2011wi, DAgnolo:2012ulg, Alonso:2013nca, Banks:2025baf}.

\section{The minimality of $SU(2)_q \times U(1)_X$}
\label{sec:SU(2)}

The smallest continuous subgroup of $ U(3)^5 $, a single $ U(1) $, can address the SM flavor puzzle via the classic Froggatt--Nielsen (FN) mechanism~\cite{Froggatt:1978nt, Leurer:1992wg, Leurer:1993gy, Fedele:2020fvh}. This $U(1)$ may be viewed as a particular linear combination of the Cartan generators, giving fermions family-dependent charges~\footnote{The Cartan of $U(3)^5$ has rank~15, corresponding to $U(1)^{15}$.}. However, the resulting selection rules on SMEFT operators typically induce large FCNC, leading to bounds far above the TeV scale; see, e.g.,~\cite{Lalak:2010bk, Cornella:2023zme, Cornella:2024jaw}.

The Supplemental Material classifies the Lie algebras of connected compact groups with fewer than four generators and excludes the Abelian and single-$SU(2)$ possibilities. The minimal non-Abelian protecting Lie algebra is $\mathfrak{su}(2)_q\oplus\mathfrak{u}(1)$. Here is the essence of the proof. Consider a four-fermion operator with left-handed quarks,
$ \mathcal{L}\supset \mathcal C_{ijk\ell} \Lambda_{\rm eff}^{-2}\,(\bar q_i\gamma_\mu q_j)(\bar q_k\gamma^\mu q_\ell) $.
To avoid unsuppressed contributions to $K - \bar K$ and $D - \bar D$ mixing, one must assign different charges to the first two families, $X_{q_1}\neq X_{q_2}$. This predicts $\mathcal{C}_{1111}$, $\mathcal{C}_{1122}$, $\mathcal{C}_{1221}$, and $\mathcal{C}_{2222}$ to be uncorrelated $\mathcal{O}(1)$ numbers. The rotation from the interaction to the mass basis must reproduce the Cabibbo angle, whether it is generated in the up sector, the down sector, or shared between them. This choice determines whether $K$- or $D$-mixing sets the dominant bound, but it cannot evade both. Indeed, as shown in Table~6 of Ref.~\cite{Silvestrini:2018dos}, all these $\mathcal{O}(1)$ structures are constrained at the level $\Lambda_{\rm eff}\gtrsim 300~\text{TeV}$, with very little sensitivity to the CKM alignment. In the shared case, the bound is stronger by up to a factor of $\sim 10$, due to the possibility of a 1–2 rotation phase~\cite{Blum:2009sk}. 

Multiple $U(1)$ factors also fail for the same reason. Thus, achieving a \textit{structured} cancellation among operators with $\Lambda_{\rm eff}=\mathcal{O}(\text{TeV})$ requires an $SU(2)_{q}$ flavor symmetry under which $ q \equiv (q_1,q_2)^T$ transforms as a doublet. For $SU(2)_q$-invariant operators, the Cabibbo rotation becomes unphysical and flavor violation proceeds only through the third family, as in MFV.

For the remaining SM fermions, it is not possible to charge them under the same $SU(2)$ factor and provide sufficient flavor protection due to unavoidable flavor transfers. Therefore, one needs an additional symmetry: a \emph{single} additional $U(1)_{X}$ taken as an appropriate linear combination of the remaining Cartan generators, $ U(1)_{X}\subset U(1)^{14} \subset U(3)^5 $, is sufficient. This fixes the minimal Lie algebra, $\mathfrak{su}(2)_q\oplus\mathfrak{u}(1)_X$, but neither the global form of the group nor the embedding of $U(1)_X$ into $U(1)^{14}$, which we constrain next.

\begin{table*}[t]
\centering
\renewcommand{\arraystretch}{1.4} 
\setlength{\tabcolsep}{3pt}      
\begin{tabular}{c|c|c|c|c|c|c|c}
\textbf{SM fields} & $d_1,\,u_1$ & $d_2$ & $u_2$ & $d_3$ & $q_3,\,u_3$ & $\ell_i$ & $e_i$ \\
\hline
$U(1)_X$ charge 
& $-X_{V} - 2 X_z$ 
& $X_{V}+X_z$ 
& $-X_W$ 
& $X_z-X_U$ 
& $2 X_z - X_U$ 
& $X_i$ 
& $X_i - (4 - i) X_z$ \\
\end{tabular}
\caption{Assignment of \textit{non-zero} $U(1)_X$ charges to the SM fields. See \cref{sec:spurion} for details.
}
\label{tab:charges}
\end{table*}

Interestingly, a single $SU(2)\times U(1)$ flavor symmetry~\cite{Barbieri:1995uv, Pomarol:1995xc, Barbieri:1997tu} and its recent revivals~\cite{Antusch:2023shi, Linster:2018avp, Greljo:2023bix, Greljo:2024zrj, Antusch:2025xrs, Darme:2023nsy, Calibbi:2025rxn} offer an elegant minimal solution to the SM flavor puzzle. 
However, all $ SU(2) \times U(1)$ constructions proposed so far predict sizable FCNC, see Fig.~1 of Ref.~\cite{Antusch:2023shi}. These findings, together with the established paradigms of $U(2)^5$ and MFV, make our result somewhat unexpected.

\section{Minimal setup for SM and BSM flavor puzzles} 
\label{sec:spurion}

Let us impose the identified minimal flavor group
\begin{equation}
    \mathcal{H}_{\rm MFP} = SU(2)_{q} \times U(1)_{X}\,.
\end{equation}
As shown in the Supplemental Material, under the stated assumptions the fermion embedding under $SU(2)_q$ that avoids dangerous flavor transfers is unique: only the two light left-handed quark fields transform nontrivially, forming a doublet $q \equiv (q_1,q_2)^T \sim \mathbf{2}$, while all other fields are $SU(2)_q$ singlets.

Symmetry-breaking spurions must be $SU(2)_{q}$ doublets or singlets, since higher-spin irreps lead to sizeable flavor violation~\cite{Greljo:2026pkz}. At least three doublets are needed to reproduce the CKM matrix, including its CP-violating phase, as shown in the Supplemental Material.
Therefore, the minimal symmetry-breaking spurion sector is
\begin{equation}
\boldsymbol{V}\sim \boldsymbol{2}_{X_V}, \quad  \boldsymbol{W} \sim \boldsymbol{2}_{X_W}, \quad \boldsymbol{U}\sim \boldsymbol{2}_{X_U}, \quad \boldsymbol{z}\sim \boldsymbol{1}_{X_z},
\end{equation}
where the singlet is needed, in particular, for the charged-lepton mass hierarchies.

These spurions enter the effective Lagrangian only through positive integer powers respecting symmetry covariance. Since doublet spurions will play different roles, we require $|X_V| \neq |X_W| \neq |X_U|$. All $U(1)_X$ charges are taken to be integers, without loss of generality.

We next require that Yukawa textures, built through spurion insertions, are such that $i)$ interaction-to-mass–basis rotations are sufficiently suppressed, avoiding excessive FCNC from flavor non-universal operators, and $ii)$ the observed SM flavor hierarchies are reproduced from powers of spurions of the same order. These two conditions strongly constrain the admissible textures; a simple realization is the charge assignment of \cref{tab:charges}~\footnote{The term in the up sector $\overline{q}\, \boldsymbol{U}(\boldsymbol{z}^*)^2\, u_3$ is omitted; the CKM $V_{3i}$ elements are mainly generated in the down sector. Alternatively, one could have written two terms $\overline{q}\, \boldsymbol{U z} \,d_3$ and $\overline{q}\, \boldsymbol{U} \,u_3$, in which case the charges of $q_3, u_3$ become $\eminus X_U$ while that of $d_3$ is $\eminus (X_U + X_z)$. In this case, both terms contribute similarly to the CKM elements $V_{td}$ and $V_{ts}$.}.

As a result of \cref{tab:charges}, the leading terms in the quark Yukawa Lagrangian are
\begin{align}\label{eq:Yukawa}
-\mathcal{L} &\supset
y^{u}_{3}\,\bar{q}_3\,u_3\,\widetilde{H}
+ y^{u}_{2}\,\bar{q}\,\boldsymbol{W}\,u_2\,\widetilde{H}
+ y^{u}_{1}\,\bar{q}\,\boldsymbol{V} \boldsymbol{z}^2\,u_1\,\widetilde{H} \nonumber\\
& + y^{d}_{3}\,\bar{q}_3\,\boldsymbol{z}\,d_3\,H
+ y^{d}_{2}\,\bar{q}\,\widetilde{\boldsymbol{V}}\boldsymbol{z}^*\,d_2\,H
+ y^{d}_{1}\,\bar{q}\,\boldsymbol{V} \boldsymbol{z}^2\,d_1\,H\nonumber\\
& + y^{d}_{q}\,\bar{q} \,\boldsymbol{U} \boldsymbol{z}^*\,d_3\,H+\mathrm{h.c.} \,,
\end{align}
where $\widetilde H = i \sigma_2 H^*$ and $\widetilde{\boldsymbol V}  = i \sigma_2 \boldsymbol V^*$. With an appropriate choice of $X_i$ from \cref{tab:charges} discussed later, the lepton Yukawa Lagrangian is
\begin{equation}\label{eq:YukLept}
    - \mathcal{L} \supset
y^{e}_{3}\,\bar{\ell}_3\,\boldsymbol{z}\,e_3 H + y^{e}_{2}\,\bar{\ell}_2\,\boldsymbol{z}^2\,e_2 H
+ y^{e}_{1}\,\bar{\ell}_1\,\boldsymbol{z}^3\,e_1 H + \mathrm{h.c.}\,.
\end{equation}
All $y^{u,d,e}_{x}$ are assumed to be $\mathcal{O}(1)$ parameters, while hierarchies are fully encoded in the spurions $\boldsymbol{V}$, $\boldsymbol{W}$, $\boldsymbol{U}$, and $\boldsymbol{z}$, which acquire small background values. Schematically,
\begin{equation}\label{eq:quarkY}
Y_u =
\begin{pmatrix}
\boldsymbol{V}\boldsymbol{z}^2  & \boldsymbol{W} & 0 \\
0 & 0 & 1
\end{pmatrix}\,,
\quad
Y_d =
\begin{pmatrix}
\boldsymbol{V}\boldsymbol{z}^2 & \widetilde{\boldsymbol{V}} \boldsymbol{z}^* & \boldsymbol{U} \boldsymbol{z}^* \\
0 & 0 & \boldsymbol{z}
\end{pmatrix}\,,
\end{equation}
and
\begin{equation}\label{eq:leptonY}
Y_e = \diag (\boldsymbol{z}^3, \boldsymbol{z}^2, \boldsymbol{z})\,.
\end{equation}

Exploiting $\mathcal{H}_{\rm MFP}$ transformations, as well as other global phases broken by \cref{eq:Yukawa} and \cref{eq:YukLept}, we may, \textit{without} loss of generality, represent the breaking as
\begin{equation}\label{eq:4}
\small
\boldsymbol{z} =
z,~
\boldsymbol{V} =
v\begin{pmatrix}
1   \\
0
\end{pmatrix},
~
\boldsymbol{W} =
w \begin{pmatrix}
\sin \alpha \\
\cos \alpha
\end{pmatrix},
~
\boldsymbol{U} =
u \begin{pmatrix}
\sin \beta \, e^{i \phi}\\
\cos \beta
\end{pmatrix}\,,
\end{equation}
where $ z, v, w, u , \alpha, \beta, \phi$, as well as all $y^{u,d,e}_{x}$, are real parameters. At this level, a single CP phase remains.

The Yukawa matrices admit a perturbative singular value decomposition,
$Y_f = L_f \hat Y_f R_f^\dagger$,
where $L_f$ and $R_f$ rotate from the mass to the interaction basis and
$\hat Y_f$ contains the diagonal Yukawas. In the SM, only the CKM matrix,
$V_{\rm CKM} = L_u^\dagger L_d$,
is observable, while all rotations become physical in the SMEFT. 

Remarkably, spurions of comparable magnitude, $z,v,w,u\sim\mathcal{O}(10^{-2})$, with all other parameters of $\mathcal{O}(1)$, fit the SM fermion masses and mixings, successfully reproducing the observed flavor hierarchies. As detailed in the Supplemental Material, $z, v, w$ set the mass hierarchies, $u$ fixes $V_{ts}$, the Cabibbo angle is $\approx \eminus \alpha$, and $\beta\,e^{i\phi}\approx V_{td}^*/V_{ts}^*$.

\textbf{Alignment.} The key feature is the prediction of parametrically small
rotation angles in $L_f$ and $R_f$ while reproducing the CKM matrix. In particular, the $1\text{--}2$ block of
$Y_d$ is exactly diagonal, implying that the corresponding mixing angles arise only from products of third-generation mixings and are therefore highly suppressed, naturally evading stringent constraints from
$K\text{--}\bar K$ mixing. The Cabibbo angle is generated entirely in the up
sector, and since $Y_u^{21}=0$, this predicts a right-handed up-quark $1\text{--}2$ mixing angle $\theta^{12}_{u_R}=\lambda\, m_u/m_c\sim5\times10^{-4}$. This leads
to an irreducible contribution to $D\text{--}\bar D$ mixing, which remains safely below current bounds.

Depending on the $U(1)_X$ charge assignments, higher-order spurion insertions can
perturb the exact textures of \cref{eq:quarkY,eq:leptonY}, potentially inducing
larger mixings. Requiring sufficient alignment between interaction and mass
bases, therefore, imposes conditions on the $U(1)_X$ charges listed in the Supplemental Material. For example, the off-diagonal terms in \cref{eq:leptonY} vanish to all orders in insertions of $\boldsymbol{z}$ when 
$(X_i - X_j) \bmod X_z \neq 0\,, \forall\, i \neq j$. This assumption can be relaxed, but is adopted for simplicity. The sufficient condition $|X_a|+|X_b|<X_z$ for every $a,b\in\{U,V,W\}$ suppresses unwanted doublet insertions and maintains the alignment.

\begin{figure*}
  \centering
  \includegraphics[trim={0cm 0cm 0cm 0cm}, clip, width=1.01\textwidth]{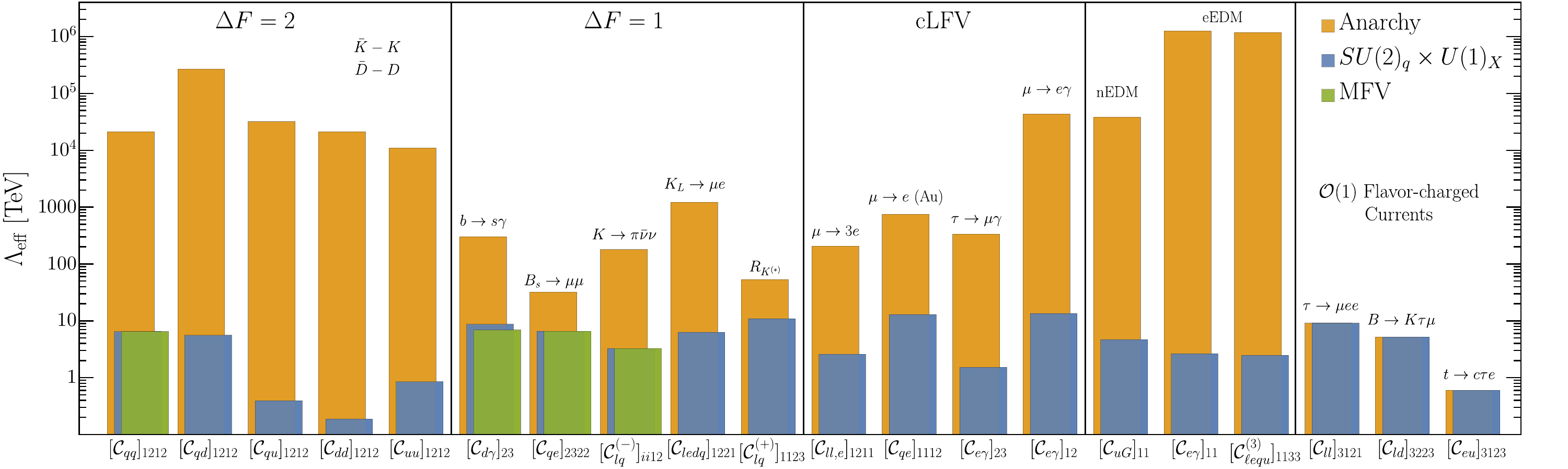}
  \caption{Bounds on selected SMEFT operators: Anarchy (yellow), $SU(2)_{q} \times U(1)_{X}$ (blue), MFV (green). See Sec.~\ref{sec:SMEFT} for details.}
  \label{fig:bounds}
\end{figure*}

\vspace{0.2cm}
\noindent

\textbf{Spurion insertions.} Beyond suppressed FCNC from small rotations of $\Delta F=0$ non-universal operators, one must ensure that direct spurion insertions generating $\Delta F\neq 0$ operators in the interaction basis arise only at sufficiently high order, which imposes additional constraints on the spurion charges. We outline the full set of resulting conditions in the Supplemental Material.

To work with concrete criteria, for all dimension-6 $\Delta B=0$ SMEFT operators in the Warsaw basis~\cite{Grzadkowski:2010es} we are imposing an effective scale
$\Lambda_{\rm eff}\lesssim10\,\text{TeV}$, where~\footnote{The relation between $\Lambda_\text{eff}$ and new mass thresholds is, of course, UV-dependent. Regarding dimension-8 operators~\cite{Murphy:2020rsh}, under flavor anarchy with $\mathcal{C}_8 \sim 1$, only a small subset can probe scales as high as $\Lambda_\text{eff,8} \lesssim 100$\,TeV~\cite{Liao:2024xel}. However, in our setup, spurions suppressing dimension-6 also suppress all such contributions.}
\begin{equation}\label{eq:EffectiveScale}
    \mathcal{L} \supset \frac{\mathcal C}{\Lambda^2_\text{eff}} Q\,.
\end{equation}
Here, $\mathcal C$ encodes \emph{only} the flavor spurion power counting. For dipole operators, we do not assume a loop factor but do include one gauge coupling for each field-strength tensor. The numerical values of the $U(1)_X$ charges consistent with \cref{tab:charges} and the aforementioned conditions are not unique, in a spirit similar to the FN setup.

\section{SMEFT analysis for a  selected benchmark} 
\label{sec:SMEFT}
In this section, we present a representative benchmark:
\begin{align}\label{eq:chargesBench}
    \{X_{V},\, X_{W},\, X_{U},\, X_{z}\} 
 &= \{1,\,\eminus2,\,\eminus3,\,7\}\,, \nonumber\\
  \{X_1,\, X_2,\, X_3\} &= \{\eminus2,\,\eminus3,\,\eminus1\}\,.
\end{align}
Following the methodology~\cite{Greljo:2022cah}, we systematically construct the basis of operators order by order in the spurion expansion.

In \cref{tab:SMEFT_count}, we compare the number of flavor-symmetric operator structures that require no spurion insertions, i.e.~$\mathcal{O}(1)$, in four scenarios: $U(3)^5$, $U(2)^5$, $SU(2)_q \times U(1)^{13}$, and $SU(2)_q \times U(1)_X$. The third case is the symmetry of the self-conjugate SMEFT plus the 4 non-Hermitian operators with the same structure as the top Yukawa~\footnote{The group is $U(1)^{13}$ instead of $U(1)^{14}$ since the top Yukawa breaks $U(1)_{q_3} \times U(1)_{u_3} \rightarrow U(1)_{q_3 + u_3}$.}. Importantly, $SU(2)_q \times U(1)^{13}$ describes the subset of $SU(2)_q \times U(1)_X$ operators that are allowed \emph{independent of the choice of charges}, and thus the irreducible operator set for MFP phenomenology at $\mathcal{O}(1)$. This includes all flavor-conserving but universality-violating operators, a broader set than under $U(2)^5$. Indeed, \cref{tab:SMEFT_count} shows that only 5 operators in our benchmark depend on the $U(1)_X$ charge assignment. These are necessarily non-Hermitian, and we indicate them with colored boxes in the Supplemental Material, where we provide a complete basis up to quadratic order.

\begin{table}[t]

\input{Tab_SMEFT_count}
\caption{Number of independent, $\Delta B= 0$ dimension-6 SMEFT operators that are invariant under the specified flavor symmetry at $\mathcal{O}(1)$, i.e. without spurion insertions. For $SU(2)_{q} \times U(1)_{X}$, the counting assumes \cref{eq:chargesBench}.
}
\label{tab:SMEFT_count}
\end{table}

Next, we allow spurion insertions and examine the resulting flavor-changing operators individually to confirm that FCNCs are sufficiently suppressed. In constructing the operator basis, we impose only the flavor symmetry, without assuming CP invariance, and explicitly distinguish the independent CP-even and CP-odd $\mathcal{O}(1)$ coefficients. More precisely, for every independent operator at a fixed order in the spurion expansion, one applies the appropriate rotation matrices from the interaction to the mass basis and confronts the experimental bounds. We review the compendium of SMEFT bounds compiled in~\cite{Silvestrini:2018dos, Kley:2021yhn, Greljo:2022jac, Feruglio:2015gka, Davidson:2020hkf, Plakias:2023esq, Delzanno:2024ooj, Calibbi:2017uvl, Panico:2016ull, Aebischer:2020dsw, Belle:2022pcr, Brod:2022bww, Grunwald:2025kot, Aebischer:2018csl}, and present a selection of them in \cref{fig:bounds}. 

Flavor symmetry alone does not solve the flavor-blind CP problem. It neither constrains the purely bosonic operator $GG\widetilde G$~\cite{Kley:2021yhn} nor generically suppresses the electron EDM sufficiently~\cite{Alarcon:2022ero, Antusch:2023shi}. In \cref{fig:bounds} we show the irreducible contribution obtained when CP is broken only by $\boldsymbol U$, the spurion responsible for the CKM phase. More generally, write additional flavor-singlet CP breaking as $\epsilon_{\rm CP}$, so that, for example, the electron dipole contains $z^3\epsilon_{\rm CP}[Q_{e(B,W)}]_{11}$. With the adopted normalization, TeV-scale tree- (loop-) level generated dipoles require $\epsilon_{\rm CP}\lesssim10^{-4}\,(10^{-2})$. This approximate CP assumption is additional to MFP. It is needed only for flavor-diagonal EDMs; the CP-odd FCNC coefficients included in our analysis may remain $\mathcal O(1)$ because their amplitudes already receive flavor-spurion suppression.

Finally, \cref{fig:bounds} summarizes the FCNC and EDM bounds on $\Lambda_\text{eff}$ defined in \cref{eq:EffectiveScale} for a selected set of operators and compares them with the two limiting cases of MFV and flavor anarchy. We see that $SU(2)_{q} \times U(1)_{X}$ preserves the rich pattern of flavor-changing phenomena exhibited in the anarchy case, while remaining compatible with new dynamics as light as $\Lambda_\text{eff} \lesssim 10$\,TeV. A similar limit holds for MFV, but it severely restricts the allowed flavor dynamics. Interestingly, the benchmark in~\cref{eq:chargesBench} permits the following \emph{flavor-charged currents} at $\mathcal{O}(1)$: $ [Q_{eu}]_{3123}$, $[Q_{\ell \ell}]_{3121}$, and $ [Q_{\ell d}]_{3223}$ which induce unsuppressed $t \to c \tau^- e^+$~\cite{Gottardo:2018ptv}, $\tau^- \to \mu^+ e^- e^-$~\cite{Plakias:2023esq, Bigaran:2022giz}, and $B^+ \to K^+ \tau^+ \mu^-$~\cite{Belle:2022pcr} transitions. The mediators generating these interactions are necessarily charged under $\mathcal{H}_{\rm MFP}$, which motivates the term ``flavor-charged currents". Such $\mathcal{O}(1)$ flavor-charged currents cannot arise in MFV or $U(2)^5$, where every flavor-violating structure is built from, and hence aligned with, the SM Yukawa spurions. Unlike the maximal flavor universality violation identified by $SU(2)_q \times U(1)^{13}$, they strongly depend on the specific $U(1)_X$ charges.

\section{Outlook and Conclusions} 
\label{sec:conc}

In this letter, we introduce a novel mechanism that ensures the structural protection of generic TeV-scale NP by identifying the minimal continuous flavor symmetry that must be approximately conserved. In contrast to the MFV paradigm, our \emph{Minimal Flavor Protection} framework allows for genuinely rich TeV-scale flavor dynamics while still satisfying stringent experimental constraints. As in MFV and $U(2)^5$, the flavor symmetry does not address flavor-blind CP violation: EDM bounds require approximate CP invariance, whereas CP-odd FCNC coefficients may remain $\mathcal{O}(1)$.

Our framework predicts a broad spectrum of flavor-changing and universality-violating dynamics, as well as unconventional flavor-charged currents, strongly motivating a diverse flavor physics program. In particular, our work provides a symmetry-based argument that the set of flavor-violating dynamics accessible to current and future colliders is much larger than previously understood. This provides strong support for a future high-energy physics strategy that will explore complementarity between the intensity and energy frontiers, such as the FCC program~\cite{FCC:2025lpp, deBlas:2025gyz}.

On the phenomenological side, our framework expands the space of well-motivated higher-dimensional operators relevant for collider studies and calls for a broader program to target flavor-non-universal interactions. It also enables a systematic classification of heavy mediators that couple linearly to the SM, organized as irreducible representations of the underlying flavor symmetry, analogous to the MFV analysis of Ref.~\cite{Greljo:2023adz}. Such a classification would define a comprehensive set of TeV-scale resonances to be searched for at the LHC. The same logic can be applied to structured UV scenarios, including MSSM soft terms~\cite{Dimopoulos:1995ju, Isidori:2019pae} and composite Higgs models with partial compositeness~\cite{Kaplan:1991dc, Glioti:2024hye, Stefanek:2024kds}.

The MFP framework also offers a fresh perspective on the flavor puzzle, motivating new model-building efforts to explain the dynamical origin of the approximate symmetries identified. We have verified that departing from the charge assignments given in~\cref{tab:charges} allows for gauge-anomaly-free solutions, though alignment may not be guaranteed. A systematic exploration of viable charge assignments is therefore well motivated, as it also charts the landscape of possible flavor-charged currents.

\section*{Acknowledgments}

We thank Wolfgang Altmannshofer, Stefan Antusch, Claudia Cornella, Javier Fuentes-Mart\'in, Gian F. Giudice, Stefania Gori, Gino Isidori, Claudio A. Manzari, David Marzocca, Anders E. Thomsen, Alessandro Valenti, and Luca Vecchi for helpful comments and discussions. This work has received funding from the Swiss National Science Foundation (SNF) through the Eccellenza Professorial Fellowship ``Flavor Physics at the High Energy Frontier,'' project number 186866. The work of AP is supported by MCIU/AEI/10.13039/501100011033 (grants CEX2023-001292-S and PID2023-146220NB-I00). The work of BAS is supported by a CDEIGENT grant from the Generalitat Valenciana with no. CIDEIG/2023/35.

\section*{End Matter: Towards an ultraviolet completion of the Yukawa sector} 
\label{sec:model}

The EFT construction points naturally toward a UV framework capable of explaining the origin of the observed flavor hierarchies. A simple renormalizable realization can be obtained by introducing a vector-like fermion (VLF) “chain” structure that communicates small symmetry breaking through sequential mixings. In this setup, the Yukawa textures in Eqs.~\eqref{eq:quarkY} and \eqref{eq:leptonY} arise from a \emph{softly} broken global $SU(2)_{q} \times U(1)_{X}$ symmetry. The symmetry breaking enters only through mass-mixing terms, which may, in turn, originate from the vacuum expectation values of scalar fields in a Froggatt--Nielsen-like mechanism. For clarity, we first explain Yukawa generation through mass mixings, and then comment on how flavons dynamically generate these mixings via marginal interactions and a consistent scalar potential.

\begin{figure*}[t]
    \centering
    \includegraphics[width=1\linewidth]{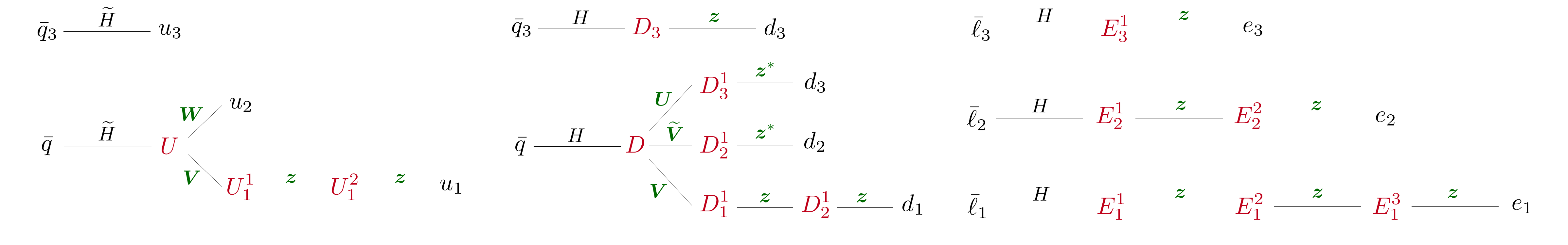}
    \caption{The chains of vector-like fermions yielding $Y_{u,d,e}$ in the TeV-scale flavor model. Flavor-spurion mixing is shown in green, and the vector-like fermions in red.}
    \label{fig:model}
\end{figure*}

The model is sketched in \cref{fig:model}. To generate the structure of $Y_e$ in Eq.~\eqref{eq:leptonY}, we introduce vector-like leptons $E_i^p$ transforming as $e_R$ under the SM gauge group. The flavor index $i=1,2,3$ labels the $e,\mu,\tau$ sectors, while $p$ denotes the position in a chain. The chains have lengths $k_i=(3,2,1)$ for $(e,\mu,\tau)$, respectively. Each $E_i^p$ carries a $U(1)_X$ charge
\begin{equation}
    X_{E_i^p} = X_i - (p-1)\,X_z\,, 
    \qquad p = 1, \ldots, k_i \, .
\end{equation}
such that only nearest-neighbor mass mixings are allowed by symmetry. Taking all VLF masses to be of order $M$, and all allowed mass mixings to be of order $M \cdot z$ with $z$ given in Eq.~\eqref{eq:4}, reproduces the charged-lepton Yukawa texture of Eq.~\eqref{eq:leptonY} after integrating out the VLFs.

The VLF content in the quark sector consists of right-handed quark partners $U$ and $D$, which form $SU(2)_q$ doublets each and are neutral under $U(1)_X$, together with the singlets $U^r_1$ and $D^r_1$ ($r=1,2$) with charges $\eminus X_V - (r-1) X_z$, as well as, $D_2^1$, $D^1_3$ and $D_3$ with charges $X_V$, $\eminus X_U$, and $2X_z - X_U$, respectively. As before, all VLF masses are of order $M$, while the allowed mass mixings are of order $M \cdot x$, with $x = z, v, w,$ or $u$ as defined in Eq.~\eqref{eq:4}. With the only CP-breaking spurion appearing in the $\bar D D_3^1$ mixing term, the quark Yukawa texture in Eq.~\eqref{eq:quarkY} follows directly, after integrating out VLFs.

The origin of VLF mass mixings could be flavor-symmetry-invariant marginal interactions with spurions promoted to dynamical scalar fields, flavons. Including soft-breaking terms in the flavon scalar potential, which are needed in any case to avoid massless Goldstone modes, we find that the minimization of the potential can naturally yield the vacuum structure of \cref{eq:4}. A detailed study of this and related constructions is left for future work.

Finally, our minimal flavor protection mechanism permits this model to be realized at the TeV scale, $M =\mathcal{O}(\mathrm{TeV})$, making this theory of flavor accessible to collider searches~\footnote{For a different TeV-scale flavor model with chains of VLFs, see~\cite{Arkani-Hamed:2026wwy}. Unlike in our case, the predicted 1–2 mixing is sizable and would lead to excessively large meson mixing in the presence of additional generic TeV-scale dynamics.}.

\vspace{0.2cm}
\noindent
\textbf{Neutrinos.} In this work, we focused on quarks and charged leptons, as the impact of tiny neutrino masses on the observables of interest is negligible. Here, we briefly comment on how neutrino flavor fits into this framework. 

To incorporate neutrino masses, we study the flavor spurion insertions in the Weinberg operator, $\frac{c_{ij}}{\Lambda} \ell_i \ell_ j H H$. When the condition
\begin{equation}
(X_i +X_j) \bmod X_z \neq 0 \quad \forall\, i,j\,, 
\end{equation}
holds, the Weinberg operator is never populated by $z$ insertions. This is indeed the case for the benchmark charge assignment in \cref{eq:chargesBench}. In such cases, the leading contributions to $c_{ij}$ arise from doublet–spurion bilinears $\boldsymbol a^\dagger \boldsymbol b$ or $\boldsymbol a^\dagger \widetilde{\boldsymbol b}$, with $\boldsymbol a,\boldsymbol b \in \{\boldsymbol V, \boldsymbol W, \boldsymbol U\}$ and $\boldsymbol a \neq \boldsymbol b$. When
\begin{equation}
|X_i +X_j| = \big| [\boldsymbol a^\dagger \boldsymbol b] ~\text{or}~[\boldsymbol a^\dagger \widetilde{\boldsymbol b}] \big| \quad \forall\, i,j \,,
\end{equation}
all entries of the Weinberg operator are populated at $\mathcal{O}(s^2)$ in the spurion expansion. As a result, we expect $c_{ij}$ to be anarchic, a welcome feature that arises because $\boldsymbol V, \boldsymbol W, \boldsymbol U$ are of comparable size.

For the benchmark charge assignment in \cref{eq:chargesBench}, all entries of $c_{ij}$ arise at $\mathcal{O}(s^2)$ except for $c_{22}$, which is additionally suppressed and first appears at $\mathcal{O}(s^3)$. Nevertheless, this texture naturally accommodates the large (apparently anarchic) PMNS mixing and the observed neutrino mass splittings. The overall scale, $\Lambda \sim \mathcal{O}(10^{11}\,\text{GeV})$, follows from fitting the mass splittings together with the cosmological upper bound on the absolute mass scale and is compatible with standard thermal leptogenesis~\cite{Davidson:2002qv}, see also~\cite{Greljo:2025suh}. To sum up, our framework straightforwardly accommodates the observed neutrino masses and mixings, in contrast with the minimally-broken $U(2)^5$ symmetry, which inevitably imprints hierarchical structures onto the PMNS matrix that must then be undone to match data.


\bibliographystyle{JHEP}
\bibliography{refs.bib}

\appendix

\clearpage
\begin{widetext}

\section*{Supplemental Material}
\renewcommand{\theequation}{S\arabic{equation}}
\setcounter{equation}{0}

\subsection{Minimality among continuous flavor symmetries}

Here we establish the minimality statement used in the main text. We consider connected compact subgroups of $U(3)^5$, linearly realized on the SM fermions, and assume independent generic $\mathcal O(1)$ coefficients for every symmetry-invariant baryon-number-conserving dimension-six SMEFT operator. Whenever a second fermion species transforms nontrivially under the same $SU(2)$, we restrict to the light-family embeddings relevant to hierarchical flavor constructions: its two lightest mass generations lie in a common nontrivial $SU(2)$ orbit. We call a symmetry viable if the strongest experimental lower bound on $\Lambda_{\rm eff}$, using the normalization of \cref{eq:EffectiveScale}, is $10\,$TeV or below. Accidental cancellations among independent invariants are not regarded as flavor protection. A connected
compact group with fewer than four continuous generators is either Abelian or,
up to global identifications, a single \(SU(2)\). It is therefore sufficient
to exclude these two possibilities.

\paragraph{Lemma 1: continuous Abelian symmetries are insufficient.}

The action of any collection of \(U(1)\) factors can be simultaneously
diagonalized. Here and below, \(i,j=1,2\) label the two light-quark
generations. Let \(\mathbf{X}_{q_i}\) denote the complete vector of Abelian charges
of \(q_i\). If
\[
    \mathbf{X}_{q_1}=\mathbf{X}_{q_2},
\]
the flavor-changing current \(\bar q_1\gamma_\mu q_2\) is neutral, so the
symmetry permits
\begin{equation}
    \bigl(\bar q_1\gamma_\mu q_2\bigr)
    \bigl(\bar q_1\gamma^\mu q_2\bigr)
    +\mathrm{h.c.},
\end{equation}
which generates unsuppressed neutral-meson mixing.

Suppose instead that \(\mathbf{X}_{q_1}\neq\mathbf{X}_{q_2}\). The off-diagonal current
may then be forbidden, but the diagonal currents
\[
    J_i^\mu\equiv \bar q_i\gamma^\mu q_i
\]
are always neutral. Operators such as
\(J_1^\mu J_{1\mu}\), \(J_2^\mu J_{2\mu}\), and
\(J_1^\mu J_{2\mu}\) are therefore allowed with unrelated coefficients.
Under a rotation
\begin{equation}
    q_1=c_\theta q'_1+s_\theta q'_2,
    \qquad
    q_2=-s_\theta q'_1+c_\theta q'_2,
\end{equation}
one finds, for example,
\begin{equation}
    J_1^\mu J_{1\mu}
    \supset
    c_\theta^2s_\theta^2
    \left[
      \bigl(\bar q'_1\gamma_\mu q'_2\bigr)^2+\mathrm{h.c.}
    \right].
\end{equation}
The left-handed rotations in the up- and down-quark sectors obey
\[
    L_u^\dagger L_d=V_{\rm CKM}.
\]
Consequently, the Cabibbo rotation cannot be absent from both sectors:
changing its location merely moves the resulting constraint between
\(D\)- and \(K\)-meson mixing. Avoiding these contributions would require
special relations among independent Wilson coefficients that no Abelian
symmetry enforces. This excludes any continuous \(U(1)^n\) symmetry,
independently of \(n\).

\paragraph{Lemma 2: a single \(SU(2)\) is insufficient.}

For three generations, an \(SU(2)\) can act on a given SM fermion species as
\[
    \mathbf{1}\oplus\mathbf{1}\oplus\mathbf{1},
    \qquad
    \mathbf{2}\oplus\mathbf{1},
    \qquad\text{or}\qquad
    \mathbf{3}.
\]
We use \(a,b=1,2,3\) for three-generation or triplet indices and
\(i,j=1,2\) for the components of an \(SU(2)\) doublet. If the three quark
weak doublets \(q_a\) are
all singlets, arbitrary flavor-changing currents are allowed. If instead
\(q_a\) forms a triplet, it transforms as \(q_a\to R_{ab}q_b\), with \(R\) a
real orthogonal matrix. This immediately permits the invariant
\begin{equation}
    \sum_{a,b=1}^{3}
    \bigl(\bar q_a\gamma_\mu q_b\bigr)
    \bigl(\bar q_a\gamma^\mu q_b\bigr)
    +\mathrm{h.c.}.
    \label{eq:minimality-triplet-dangerous}
\end{equation}
The term with \(a=1\) and \(b=2\) is
\((\bar q_1\gamma_\mu q_2)^2\), and therefore gives unsuppressed \(K\)- or
\(D\)-meson mixing. The only viable action on the quark doublets is thus
\begin{equation}
    q_i=(q_1,q_2)^T\sim\mathbf{2},
    \qquad i=1,2,
    \qquad
    q_3\sim\mathbf{1},
\end{equation}
which is the \(SU(2)_q\) identified in the main text.

Consider now the three lepton weak doublets \(\ell_a\), with \(a=1,2,3\)
a generation index. If all three are \(SU(2)_q\) singlets, the symmetry
allows arbitrary charged-lepton-flavor-violating structures, for example $\bigl(\bar\ell_1\gamma_\mu\ell_2\bigr) \bigl(\bar\ell_1\gamma^\mu\ell_1\bigr)$.
In the charged-lepton mass basis, this operator contains a contribution
to \(\mu\to3e\); analogous structures contribute to \(\mu\)-to-\(e\)
conversion.

Alternatively, suppose that the leptons transform nontrivially, either as
\(\mathbf{2}\oplus\mathbf{1}\) or as a \(\mathbf{3}\), under the same \(SU(2)\). In both
cases the symmetry permits the adjoint-current invariant
\begin{equation}
    \mathcal{O}_{q\ell}
    =
    \sum_{A=1}^{3}
    \bigl(\bar q\gamma_\mu t_q^A q\bigr)
    \bigl(\bar\ell\gamma^\mu t_\ell^A\ell\bigr),
    \label{eq:minimality-flavor-transfer}
\end{equation}
where \(A=1,2,3\) is the \(SU(2)\) adjoint index, while \(t_q^A\) and
\(t_\ell^A\) denote the generators in the quark and lepton representations,
respectively. The subscripts \(q\) and \(\ell\) are representation labels,
not summed indices. For the
\(\mathbf{2}\oplus\mathbf{1}\) case, let \(\ell_i\), with \(i=1,2\), denote the
lepton doublet. The invariant in
Eq.~\eqref{eq:minimality-flavor-transfer} then contains ``flavor-transfer" operators~\cite{Darme:2023nsy, Antusch:2023shi} of the form
\begin{equation}
    \mathcal{O}_{q\ell}
    \supset
    \frac12
    \bigl(\bar q_1\gamma_\mu q_2\bigr)
    \bigl(\bar\ell_2\gamma^\mu\ell_1\bigr)
    +\mathrm{h.c.}.
\end{equation}
In the down-quark mass basis, identifying the two lepton-doublet
components with the first and second charged-lepton generations, the
charged-lepton component contains a term of the form
\begin{equation}
    \bigl(\bar d_L\gamma_\mu s_L\bigr)
    \bigl(\bar\mu_L\gamma^\mu e_L\bigr),
\end{equation}
which directly mediates \(K_L\to\mu^\pm e^\mp\). Its very strong
experimental limit probes effective contact scales far above the TeV range.
For a lepton triplet, with components \(\ell_a\), \(a=1,2,3\), it is easy to check that the same
adjoint invariant contains analogous off-diagonal currents between triplet
components.

Together with Lemma~1, this proves that at least four
continuous generators are required. At this dimension the minimal
non-Abelian Lie algebra is
\begin{equation}
    \mathfrak{su}(2)_q\oplus\mathfrak{u}(1),
\end{equation}
and the explicit \(SU(2)_q\times U(1)_X\) construction in the main text
provides the corresponding existence proof.

\paragraph{Remark on discrete flavor symmetries.}
Our minimality statement concerns \emph{connected} flavor groups.
Instead, consider an arbitrary subgroup $\Gamma\subset U(2)_q$ act on the light-family doublet
$q=(q_1,q_2)^T$. Overall $U(1)$ phases drop out of quark bilinears. The traceless currents
$J_\mu^a=\bar q\gamma_\mu\sigma^a q$ therefore transform as a three-vector
under the induced $SO(3)$ action, using $\overline{\Gamma}\subset U(2)/U(1)_{\rm center}\simeq SO(3)$. For generic Wilson coefficients,
simultaneous protection of $K$ and $D$ mixing requires the symmetry to
enforce an isotropic contraction in this space: if a preferred axis is
left invariant, independent longitudinal and transverse contractions are
allowed and, after rotating between the non-aligned up- and down-quark
bases, generically generate
$(\bar q_1\gamma_\mu q_2)^2+\mathrm{h.c.}$.

This observation also covers disconnected extensions of the connected
Abelian candidates. A non-central $U(1)\subset U(2)_q$ acts projectively as
rotations about a fixed axis, and its disconnected completions can at most
preserve or reverse this axis. For example,
$O(2)_q=U(1)\rtimes Z_2$ allows the invariant $(J_\mu^3)^2$; after a flavor
rotation by an angle $\theta$,
\begin{equation}
    J^3_\mu\to   \cos(2\theta)\,J^{\prime 3}_\mu+
\sin(2\theta)\,J^{\prime 1}_\mu ,
\end{equation}
so $(J_\mu^3)^2$ contains
$\sin^2(2\theta)(\bar q'_1\gamma_\mu q'_2)^2+\mathrm{h.c.}$. Similarly, for the maximal torus
$T=U(1)_{q_1}\times U(1)_{q_2}$ one has
$N_{U(2)}(T)/T\simeq Z_2$, corresponding to
$q_1\leftrightarrow q_2$. Thus such extensions still allow anisotropic
flavor contractions and do not enforce simultaneous $K$- and $D$-mixing
protection.

Purely finite groups provide a genuine alternative. The finite subgroups of
$SO(3)$ are $C_n$, $D_n$, $A_4$, $S_4$, $A_5$.
The cyclic and dihedral groups preserve an axis, whereas $A_4$ is the
smallest one for which the geometric ${\bf3}$ is irreducible. Since $A_4$
has no faithful two-dimensional representation, its action on the quark
doublet is lifted to its double cover, the \textit{binary tetrahedral} group
\begin{equation}
    2T/Z_2\simeq A_4, \qquad 2T \subset SU(2), \qquad  |2T|=24,
\end{equation}
for which ${\bf2}^\ast\otimes{\bf2}={\bf1}\oplus{\bf3}$,
with the ${\bf3}$ irreducible. The flavor-adjoint indices therefore admit
a unique quadratic contraction, $\delta_{ab}J_\mu^aJ^{b\mu}$, exactly as
for $SU(2)_q$. Hence $2T$ reproduces the $SU(2)_q$ protection of the
leading spurion-independent $\Delta F=2$ four-quark operators and is the
minimal purely finite alternative.

A central $U(1)$ does not change this conclusion, since it acts as a common
phase on $q$ and trivially on $\bar q q$; e.g.\ $U(1)_{\rm c}\!\cdot 2T$
derives its flavor protection entirely from $2T$. Thus $SU(2)_q$ is
minimal among connected symmetries, while allowing finite groups leads to
$2T$ as the minimal alternative. At higher operator or spurion order,
additional $2T$ invariants can distinguish the finite theory from
continuous $SU(2)_q$.

\subsection{Fermion representations and the minimal flavor breaking sector}

\paragraph{Assumptions and scope.}
The light left-handed quarks transform as
\begin{equation}
    q=(q_1,q_2)^T\sim\mathbf 2,
    \qquad q_3\sim\mathbf 1
\end{equation}
under $SU(2)_q$. Whenever another three-generation fermion species transforms as
$\mathbf 2\oplus\mathbf 1$, we assume that its first two mass generations form the doublet and its third generation is the singlet. Nontrivial $SU(2)_q$-breaking spurions are taken to enter the Yukawa sector linearly; arbitrary powers of $SU(2)_q$ singlets such as $z$ may accompany them. The FCNC motivation for this conservative assumption is given below.

We additionally assume that, to the spurion order under consideration, the overall phase of each spurion is an independent reparametrization redundancy. A simple sufficient condition is that every right-handed fermion couples to only one independent net spurion-phase combination, or more generally that all terms involving a given right-handed field carry the same net spurion phase.

\section*{1. Only $q$ transforms nontrivially under $SU(2)_q$}

For a three-generation fermion species $f=u,d,\ell,e$, the possible restrictions to $SU(2)_q$ are
\begin{equation}
    \mathbf1\oplus\mathbf1\oplus\mathbf1,
    \qquad
    \mathbf2\oplus\mathbf1,
    \qquad
    \mathbf3.
\end{equation}
We exclude the nontrivial possibilities in turn.

\paragraph{Triplets.}
If $f_a$, $a=1,2,3$, transforms as a real triplet, the exact symmetry permits
\begin{equation}
    \sum_{a,b=1}^{3}
    \bigl(\bar f_a\gamma_\mu f_b\bigr)
    \bigl(\bar f_a\gamma^\mu f_b\bigr)
    +\mathrm{h.c.},
\end{equation}
independently of its common $U(1)_X$ charge. This contains
$(\bar f_1\gamma_\mu f_2)^2$ and hence unsuppressed same-sector flavor violation. No SM fermion species may therefore transform as a triplet.

\paragraph{Left-handed leptons.}
If $\ell=(\ell_1,\ell_2)^T\sim\mathbf2$, the charge-independent invariant
\begin{equation}
    \sum_{A=1}^{3}
    \bigl(\bar q\gamma_\mu t^Aq\bigr)
    \bigl(\bar\ell\gamma^\mu t^A\ell\bigr)
\end{equation}
Using the fundamental completeness relation
\begin{equation}
    \sum_{A=1}^{3}(t^A)_{ij}(t^A)_{kl}
    =\frac12\delta_{il}\delta_{jk}
     -\frac14\delta_{ij}\delta_{kl},
    \label{eq:fund-completeness}
\end{equation}
this contains quark--lepton flavor-transfer currents. In aligned bases these include
\begin{equation}
    \frac12
    \bigl(\bar q_1\gamma_\mu q_2\bigr)
    \bigl(\bar\ell_2\gamma^\mu\ell_1\bigr)
    +\mathrm{h.c.},
\end{equation}
which mediate $K_L\to\mu e$ and $K\to\pi\nu\bar\nu$. A relative quark--lepton rotation only redistributes the same adjoint current between flavor-changing quark processes and flavor-diagonal quark currents constrained by charged-lepton conversion. Thus
\begin{equation}
    \boxed{\ell_1,\ell_2,\ell_3\sim\mathbf1.}
\end{equation}

\paragraph{Right-handed charged leptons.}
We work in the convention of Refs.~\cite{Barbieri:1998em,Barbieri:1999pe} in which a right-handed doublet transforms as an anti-doublet, with
$\bar t^A=-(t^A)^*$. Correspondingly,
\begin{equation}
    \sum_{A=1}^{3}(t^A)_{ij}(\bar t^A)_{kl}
    =-\frac12\delta_{ik}\delta_{jl}
     +\frac14\delta_{ij}\delta_{kl},
    \label{eq:fund-antifund-completeness}
\end{equation}
which follows from Eq.~\eqref{eq:fund-completeness}. If
$e=(e_1,e_2)^T\sim\bar{\mathbf2}$, the exact invariant
\begin{equation}
    \sum_{A=1}^{3}
    \bigl(\bar q\gamma_\mu t^Aq\bigr)
    \bigl(\bar e\gamma^\mu\bar t^Ae\bigr)
\end{equation}
contains an unsuppressed $e$--$\mu$ flavor-transfer current and therefore induces $K_L\to\mu e$ or, after a relative rotation, comparably constrained $\mu$--$e$ conversion operators. Hence
\begin{equation}
    \boxed{e_1,e_2,e_3\sim\mathbf1.}
\end{equation}

\paragraph{Right-handed quarks in the $\mathbf2$--$\bar{\mathbf2}$ convention.}
Let
\begin{equation}
    q=(q_1,q_2)^T\sim\mathbf2,
    \qquad
    f=(f_1,f_2)^T\sim\bar{\mathbf2},
    \qquad f=u,d,
\end{equation}
with $f_3\sim\mathbf1$. The exact symmetry permits the $U(1)_X$-neutral operator
\begin{equation}
    \mathcal O_{qf}^{\mathrm{adj}}
    =
    \sum_{A=1}^{3}
    \bigl(\bar q\gamma_\mu t^Aq\bigr)
    \bigl(\bar f\gamma^\mu\bar t^Af\bigr).
    \label{eq:mixed-adjoint-antidoublet}
\end{equation}
Using Eq.~\eqref{eq:fund-antifund-completeness}, it already contains in the symmetry basis
\begin{equation}
    -\frac12
    \bigl(\bar q_1\gamma_\mu q_2\bigr)
    \bigl(\bar f_1\gamma^\mu f_2\bigr)
    +\mathrm{h.c.}
\end{equation}
After rotating to the mass basis, define $\Omega_f\equiv L_f^TR_f$. The coefficient of the corresponding mixed-chirality $\Delta F=2$ operator is
\begin{equation}
    C_{LR}^{12}
    =-\frac12(\Omega_f)_{11}^*(\Omega_f)_{22}.
    \label{eq:barbieri-lr-coefficient}
\end{equation}

A linearly generated light-family Yukawa block generically takes the form~\cite{Barbieri:1998em,Barbieri:1999pe}
\begin{equation}
    Y_f^{(2)}=a_f\epsilon+S_f,
    \qquad S_f^T=S_f,
\end{equation}
where $a_f\epsilon$ is the singlet and $S_f$ the triplet. The singlet alone would give degenerate singular values. A natural hierarchical realization is
\begin{equation}
    Y_f^{(2)}\simeq
    \begin{pmatrix}0&a\\-a&b\end{pmatrix},
    \qquad |a|\ll|b|,
\end{equation}
for which perturbative diagonalization gives
\begin{equation}
    \theta_L\simeq\frac{a}{b},
    \qquad
    \theta_R\simeq-\frac{a}{b},
    \qquad
    \Omega_f\simeq\mathbf1_2.
\end{equation}
Equation~\eqref{eq:barbieri-lr-coefficient} then yields
\begin{equation}
    C_{LR}^{12}=-\frac12+\mathcal O\!\left(\frac{|a|^2}{|b|^2}\right),
\end{equation}
producing an unsuppressed LR $\Delta C=2$ operator for $f=u$ or LR $\Delta S=2$ operator for $f=d$. Avoiding this requires a fine-tuned relative orientation of the independent singlet and triplet contributions rather than symmetry protection; moreover, the neutral tensor $S_fS_f^\dagger$ permits direct $\Delta F=2$ invariants already at quadratic order.

The assignment $f\sim\mathbf2$ is physically equivalent. By pseudoreality, the fixed redefinition $\widehat f=\epsilon f$ interchanges $\mathbf2$ and $\bar{\mathbf2}$ (with the same separately assigned $U(1)_X$ charge) and maps all Yukawa and current tensors into the form above without changing masses or FCNC amplitudes. We therefore conclude
\begin{equation}
    \boxed{u_1,u_2,u_3,d_1,d_2,d_3\sim\mathbf1.}
\end{equation}
Together with the lepton results, only $q=(q_1,q_2)^T$ transforms nontrivially under $SU(2)_q$.

\section*{2. Linear charged Yukawa spurions are fundamentals}

With all right-handed fermions $SU(2)_q$ singlets, a linear term $\bar q\,S\,fH$ is invariant only if $\bar{\mathbf2}\otimes R_S$ contains a singlet. Hence
\begin{equation}
    \boxed{S\sim\mathbf2.}
\end{equation}
Thus every nontrivial $SU(2)_q$-breaking spurion entering the Yukawa sector linearly is a fundamental; $z$ remains an $SU(2)_q$ singlet.

\paragraph{FCNC motivation for linear fundamental spurions.}
Define the adjoint quark current
\begin{equation}
    J_\mu^A=\bar q\gamma_\mu t^Aq,
    \qquad A=1,2,3.
\end{equation}
The dangerous structure $(\bar q_1\gamma_\mu q_2)^2$ is the highest-weight component of the spin-two, symmetric-traceless part of $J_\mu^AJ^{B\mu}$, so its coefficient must furnish an $SU(2)_q$ spin-two tensor. For a fundamental spurion,
\begin{equation}
    S^\dagger\otimes S=\mathbf1\oplus\mathbf3,
\end{equation}
so spin two first appears with four insertions, schematically
\begin{equation}
    \left[(S^\dagger t^AS)(S^\dagger t^BS)\right]_{\mathrm{ST}}.
\end{equation}
By contrast, for any irrep of isospin $j\geq1$,
\begin{equation}
    j\otimes j^*=0\oplus1\oplus\cdots\oplus2j
\end{equation}
already contains spin two, allowing the same FCNC tensor at quadratic order; an elementary spin-two spurion can generate it linearly.

Linearity is also the most conservative power counting. If a fixed Yukawa suppression is generated by several powers of a charged spurion, the elementary breaking parameter must be larger than in a linear realization, while the first spin-two FCNC tensor is still determined by the elementary representation: fourth order for fundamentals and generically second order for $j\geq1$. Nonlinear Yukawa insertions therefore do not postpone, and generally enhance, the associated flavor violation, as shown in~\cite{Greljo:2026pkz}. Linear fundamentals are consequently the minimally breaking and maximally protective choice for generic Wilson coefficients.

\section*{3. Three fundamental doublets are minimally required for CP violation}

Consider two doublet spurions $V,W\sim\mathbf2$ and one singlet spurion $z\sim\mathbf1$. By assumption, their overall phases are independent reparametrization redundancies,
\begin{equation}
    V\to e^{i\alpha}V,
    \qquad
    W\to e^{i\beta}W,
    \qquad
    z\to e^{i\gamma}z.
\end{equation}
All $SU(2)_q$-invariant scalar polynomials can be reduced to the real norms
\begin{equation}
    V^\dagger V,
    \qquad
    W^\dagger W,
    \qquad
    |z|^2,
\end{equation}
and the potentially complex contractions
\begin{equation}
    x\equiv V^\dagger W,
    \qquad
    y\equiv V^T\epsilon W,
    \qquad
    z.
\end{equation}
Under the independent rephasings,
\begin{equation}
    x\to e^{i(\beta-\alpha)}x,
    \qquad
    y\to e^{i(\alpha+\beta)}y,
    \qquad
    z\to e^{i\gamma}z.
\end{equation}
Consider a general potentially complex monomial
\begin{equation}
    M=x^m(x^*)^n y^p(y^*)^r z^k(z^*)^\ell,
\end{equation}
where all exponents are non-negative integers. Invariance under $\alpha$, $\beta$, and $\gamma$ requires
\begin{align}
    -(m-n)+(p-r)&=0,\\
     (m-n)+(p-r)&=0,\\
     k-\ell&=0.
\end{align}
Hence
\begin{equation}
    m=n,
    \qquad
    p=r,
    \qquad
    k=\ell.
\end{equation}
Every rephasing-invariant monomial is therefore real and proportional to
\begin{equation}
    |V^\dagger W|^{2m}
    |V^T\epsilon W|^{2p}
    |z|^{2k},
\end{equation}
possibly multiplied by powers of $V^\dagger V$ and $W^\dagger W$. Imposing $U(1)_X$ neutrality can only remove further monomials; it cannot generate a CP-odd one. Thus two fundamental doublets, even together with the complex singlet $z$, cannot furnish a CP-odd rephasing invariant.

With a third doublet $U\sim\mathbf2$, however,
\begin{equation}
    \mathcal P_{VWU}
    =(V^\dagger W)(W^\dagger U)(U^\dagger V)
\end{equation}
is invariant under all independent rephasings and automatically neutral under $U(1)_X$ for arbitrary $X_V,X_W,X_U$. Consequently
\begin{equation}
    \mathcal J_{VWU}
    =\operatorname{Im}\mathcal P_{VWU}
\end{equation}
can be nonzero. Therefore three fundamental $SU(2)_q$-doublet spurions are minimally required
to obtain an irreducible CP-violating phase under the stated assumptions.

In summary, under our assumptions we have shown that generic FCNC protection and hierarchical Yukawas require
\begin{equation}
\begin{gathered}
    \Longrightarrow
    \quad
    q=(q_1,q_2)^T\sim\mathbf2,
    \quad
    u,d,\ell,e\sim\mathbf1\oplus\mathbf1\oplus\mathbf1,
    \\
    \Longrightarrow
    \quad
    \text{all linear }SU(2)_q\text{-breaking Yukawa spurions are }\mathbf2,
    \\
    \Longrightarrow
    \quad
    \text{two doublets plus }z\text{ cannot violate CP, whereas three can.}
\end{gathered}
\end{equation}
\subsection{Producing SM parameters: Singular value decomposition}

We perform a perturbative singular value decomposition, $Y_f = L_f \hat Y_f R_f^\dagger$, to reproduce the observed fermion masses and CKM mixing $V_{\rm CKM} = L_u^\dagger L_d$, thereby fixing the spurion values and confirming that all $y_x^{u,d,e}$ are indeed $\mathcal{O}(1)$. All quark and charged lepton masses are controlled by the $z$, $v$ and $w$ spurions:
\begin{equation}
    \begin{alignedat}{4}
        y_{u} &\sim v z^2\,,&\qquad 
        y_{c} &\sim w\,,&\qquad
        y_t &\sim 1\,,
        \\
        y_{d} &\sim v z^2\,,&\qquad
        y_{s} &\sim v z\,,&\qquad
        y_b &\sim z\,,
        \\
        y_{e} &\sim z^3\,,&\qquad
        y_{\mu} &\sim z^2\,,&\qquad
        y_\tau &\sim z\,.
    \end{alignedat}
\end{equation}
To reproduce the CKM matrix, we match the Wolfenstein parameterization at leading order~\cite{Wolfenstein:1983yz}
\begin{equation}
    \alpha \approx - \lambda\,,~ \tan \phi \approx \frac{\eta}{1-\rho}\,,~ \beta \approx - \lambda\, \frac{1-\rho}{\cos \phi}\,, ~  \frac{y_q^d \,u}{y^d_3} \approx A \lambda^2 \,,
\end{equation}
where we set $\lambda = 0.225$, $A = 0.826$, $\rho = 0.163$ and $ \eta= 0.357$ following PDG~\cite{ParticleDataGroup:2024cfk}. A good fit to these and observed $y_f$, taken from Eq.~(A8) in~\cite{Greljo:2023bix}, is obtained for
\begin{equation}\label{eq:spurionFit}
    z= 0.02\,,\;\;
    v = 0.015\,,\;\; 
    w = 0.006\,,\;\;
    u = 0.04 \,.
\end{equation}
For this choice of spurions, remarkably, all nine $y^{u,d,e}_{x}$ are $\mathcal{O}(1)$ with the ratio of the largest to the smallest $ \lesssim 5$. As advertised, the spurions lie within at most one order of magnitude of each other, providing a solid starting point for dynamically addressing the SM flavor puzzle.

For transparency, using the running Yukawas quoted in Ref.~\cite{Greljo:2023bix} at $\mu=10^3\,$TeV gives the coefficient-level benchmark
\begin{equation}
\begin{aligned}
(y_1^u,y_2^u,y_3^u)&=(0.76,\,0.38,\,0.67),\\
(y_1^d,y_2^d,y_3^d)&=(1.66,\,0.66,\,0.50),\\
(y_1^e,y_2^e,y_3^e)&=(0.36,\,1.51,\,0.52),
\end{aligned}
\end{equation}
so the largest-to-smallest ratio is $4.6$.

Let us discuss the rotation matrices $L_f$ and $R_f$. Since the Cabibbo angle comes entirely from the up sector in our framework, the corresponding left-handed up-quark mixing is $\theta^{12}_{u_L} \approx \lambda$. As discussed above, $SU(2)_q$ prevents any contribution of $\theta^{12}_{u_L}$ to $D$ mixing. However, the right-handed up-quark 1-2 mixing, $\theta^{12}_{u_R} = \lambda \, m_u/m_c \sim 5\times 10^{-4}$, is an unavoidable feature of this construction. Nevertheless, when applied to a non-universal operator, it still suffices to pass limits from $D \!-\!\bar D$ mixing. The left-handed mixing angles in the down sector are $(\theta^{13}_{d_L}, \theta^{23}_{d_L})\approx (V_{td}, V_{ts})$ while the right-handed counterparts $(\theta^{13}_{d_R}, \theta^{23}_{d_R})\approx (V_{td} \frac{m_d}{m_b}, V_{ts} \frac{m_s}{m_b} )$. Down-sector 1-2 mixing arises only as $\theta^{12}_{d_L}\sim \theta^{13}_{d_L} \theta^{23}_{d_L} \sim V_{td}V_{ts}$, with $\theta^{12}_{d_R}$ further suppressed by $m_d/m_s$. This provides sufficient alignment to avoid excessive contributions to $K \!-\!\bar K$ mixing from non-universal operators. Down-alignment in the $1-2$ sector was the main motivation to choose $\boldsymbol{V}$ instead of $\boldsymbol{W}$ in the second column of $Y_d$. The latter option, assigning a spurion doublet to each family, is perhaps better UV-motivated but could require some alignment in the down-sector.

\subsection{Conditions on $U(1)_X$ charges}

The rotation angles discussed above correspond to the exact structure of the Yukawa matrices presented in the main text. However, depending on the $U(1)_X$ charges, higher spurion insertions may spoil the alignment and lead to excessively large rotation angles. Following Ref.~\cite{Silvestrini:2018dos}, starting from a flavor non-universal four-fermion operator, neutral-meson mixing imposes stringent bounds on the rotation angles listed above. A careful inspection shows that these bounds require forbidding higher-order spurion insertions: specifically, $Y^{21}_{d,u}$ must be absent up to fourth order, while $Y^{12}_d$, $Y^{31}_d$, and $Y^{32}_d$ must be absent up to third order. This translates into a set of inequalities on the spurion charges, further listed below. 

Concerning lepton Yukawas, if
\begin{equation}\label{eq:cLFVcondt}
(X_i - X_j) \bmod X_z \neq 0 \quad \forall\, i \neq j\,,
\end{equation}
then off-diagonal terms in $Y_e$ vanish to all orders in $\boldsymbol{z}$ (neglecting all other spurions for the moment), perfectly aligning the mass and interaction bases. This condition may be relaxed, but we choose to adopt it such that 2-fermion SMEFT operators preserve an accidental $U(1)_e \times U(1)_\mu \times U(1)_\tau$ symmetry to all orders in $\boldsymbol{z}$. Furthermore, 4-fermion operators can violate this symmetry only for non-identical and flavor-violating currents. Examples are $[Q_{\ell e}]_{1212}$, or when all three flavors participate simultaneously, e.g. $[Q_{\ell\ell}]_{2123}$. As a result, the most dangerous processes such as $\mu \to 3e, e\gamma$ from $\boldsymbol{z}$ insertions are absent.

The leading off-diagonal contributions to $Y_e$ thus arise from the doublet spurions. The $SU(2)$ structure implies that singlets may only be constructed starting at second order as $ \boldsymbol a^\dagger  \boldsymbol b $ or $ \boldsymbol a^\dagger \widetilde{\boldsymbol b} $ ($\boldsymbol a, \boldsymbol b =\boldsymbol V, \boldsymbol W, \boldsymbol U$ and $\boldsymbol a \neq \boldsymbol b$). Their charge assignments, $\pm (X_a \pm X_b)$, should be chosen such that the $Y_e^{12,21}$ elements are suppressed at least to quartic order to ensure enough alignment between the mass and the interaction basis and, therefore, avoid bounds from $\mu \to e$ sector reported in Table\,V of~\cite{Calibbi:2017uvl}. No further constraints are needed for $Y_e^{a3}$ and $Y_e^{3a}$ ($a=1,2$) if \cref{eq:cLFVcondt} is assumed.

In addition to suppressed FCNC from small rotations applied to $\Delta F = 0$ operators, one must also verify that direct spurion insertions leading to $\Delta F \neq 0$ EFT operators in the interaction basis appear only at sufficiently high order. By systematically examining the most stringent flavor observables, one can derive a set of inequalities that constrain the allowed charge assignments. For instance, operators like $[Q_{qd}]_{1212}$ must be forbidden up to quartic order in spurion insertions to adequately suppress Kaon mixing.

Here we list the full set of conditions on $U(1)_X$ charges. We fix $X_z>0$ and take all charges to be primitive integers. Every displayed $\pm$ sign is tested independently, and every condition is imposed for all stated index choices. For compactness, define the minimum number of singlet-spurion insertions needed to neutralize a charge $Q$ by
\begin{equation}
n_z(Q)\equiv
\begin{cases}
|Q/X_z|, & Q/X_z\in\mathbb Z,\\
\infty, & Q/X_z\notin\mathbb Z.
\end{cases}
\label{eq:nzdefinition}
\end{equation}
This notation removes any ambiguity about sign choices or noninteger charge ratios.
\begin{enumerate}
    \item\textbf{Alignment conditions for quark and charged lepton Yukawas}
    \begin{enumerate}
    \item Demanding a zero in $Y_{d}^{31}$ up to third order leads to $X_U - 4 X_z - X_{V} \notin \{0, \pm X_z, \pm 2 X_z, \pm X_a \pm X_b\}$ for all distinct $a,b\in\{U,V,W\}$. Note that $\boldsymbol a^\dagger \boldsymbol b$ and $\boldsymbol a^\dagger \widetilde{\boldsymbol b}$ form $SU(2)_q$ singlets.
    \item Demanding a zero in $Y_{d}^{32}$ up to third order leads to $X_U - X_z + X_{V} \notin \{0, \pm X_z, \pm 2 X_z, \pm X_a \pm X_b\}$ for all distinct $a,b\in\{U,V,W\}$.

    \item Demanding a zero in $Y_{d}^{12}$ up to third order leads to $X_V + X_z \notin \{0,\pm X_a,\pm X_a\pm X_z\}$ for all $a\in\{U,V,W\}$, excluding the desired $\widetilde{\boldsymbol V}\boldsymbol z^*$ term.

    \item Demanding a zero in $Y_{d,u}^{21}$ up to fourth order gives $X_V+2X_z\notin\{0,\pm X_a,\pm X_a\pm X_z,\pm X_a\pm2X_z,\pm X_a\pm X_b\pm X_c\}$ for all $a,b,c\in\{U,V,W\}$ with $a\neq b$, excluding the leading $\boldsymbol V\boldsymbol z^2$ direction whose second component vanishes on the background in \cref{eq:4}.

    \item Demanding that $Y_e^{12,21}$ first appear no earlier than fourth order requires each of $X_2-X_1-2X_z$ and $X_1-X_2-3X_z$ to avoid
    \begin{equation}
    \{0,\ \pm nX_z\ (n=1,2,3),\ \pm X_a\pm X_b\pm nX_z\ (n=0,1)\}
    \end{equation}
    for all distinct $a,b\in\{U,V,W\}$. This is the charge set of $SU(2)_q$-singlet spurion monomials below fourth order; a quartic entry is sufficiently suppressed.
    \end{enumerate}
    \item \textbf{Conditions for $\boldsymbol{\Delta F = 2}$ spurion insertions}
    \begin{enumerate}
    \item For $[Q_{qd}]_{1212}$ and $[Q_{qu}]_{1212}$ to be forbidden at quadratic order in spurions,
    \begin{align}
    |X_a \pm X_b| \neq |2X_V + 3X_z|\,,
    \qquad
    |X_a \pm X_b| \neq |X_V +2X_z-X_W|\,,
    \end{align}
    must hold for all $a,b\in\{U,V,W\}$ and both sign choices.
    \item The $[Q_{dd}]_{1212}$ and $[Q_{uu}]_{1212}$ structures must first be allowed only above third order in $\boldsymbol z$, namely
    \begin{align}
    n_z(6X_z+4X_V)>3\,,
    \qquad
    n_z(4X_z+2X_V-2X_W)>3\,.
    \end{align}
    \item For $[Q_{qd}]_{1212}$ and $[Q_{qu}]_{1212}$ with a doublet-spurion pair, the first allowed monomial must satisfy, for all $a,b\in\{U,V,W\}$ and $\sigma,\tau=\pm1$,
    \begin{align}
    n_z(3X_z+2X_V+\sigma X_a+\tau X_b)>2\,,
    \qquad
    n_z(2X_z+X_V-X_W+\sigma X_a+\tau X_b)>1\,.
    \end{align}
    The first inequality applies to $Q_{qd}$ and the second to $Q_{qu}$; $n_z=\infty$ automatically passes.
    \end{enumerate}
\item \textbf{Conditions for Semi-leptonic $\boldsymbol{\mu-e}$ LFV}
\begin{enumerate}
 \item To forbid the semileptonic $\mu-e$ LFV operators such as $[Q_{\ell d, ed}]_{12st}$ from appearing at $\cO(1)$, both possible lepton-current charges must avoid every relevant quark-current charge:
\begin{equation}
\{|X_2-X_1+X_z|,|X_2-X_1|\}
\cap
\{|2X_V+3X_z|,|X_U+X_V|,|3X_z-X_U+X_V|\}
=\varnothing\,.
\end{equation}
Similarly, the operator $[Q_{\ell edq}]_{12s3,21s3}$ is subject to analogous $\mathcal O(1)$ conditions. We find
\begin{equation}
	0\neq
	\begin{cases}
		X_r + r X_z -X_p-X_U+X_V\,, &s=1
		\\
		X_r - (3-r)X_z -X_p-X_U-X_V\,, &s=2
		\\
		X_r - (3-r)X_z -X_p\,, &s=3
	\end{cases}\,,
\end{equation}
    where $(p,r)=(1,2)$ and $(2,1)$ are both tested. In the case of $K_L \rightarrow \mu e$, $[Q_{\ell d,ed}]_{1212}$ must also be absent at linear order. Equivalently, for $\delta\in\{X_2-X_1+X_z,X_2-X_1\}$ and $\sigma=\pm1$,
    \begin{equation}
        n_z(3X_z+2X_V+\sigma\delta)>1\,.
    \end{equation}
    
    \item Linear realizations of the $\boldsymbol V$, $\boldsymbol U$ and $\boldsymbol W$ spurions in operators of the form $[Q_{\ell edq}]_{prst}$ must be eliminated, since choices such as $(prst) = (1212)$, along with their permutations, generate contributions to $K_L \to \mu e$. The conditions read
    \begin{equation}
	0\neq
	\begin{cases}
		\sigma X_a + X_V + X_r - (2-r)X_z -X_p\,, \qquad &s=1
		\\
		\sigma X_a - X_V + X_r - (5-r)X_z -X_p\,, \qquad &s=2
	\end{cases}~\,,
    \end{equation}
    where $a\in\{V,U,W\}$, $\sigma=\pm1$, and both $(p,r)=(1,2),(2,1)$ are tested. Quadratic terms of the form $\boldsymbol V\boldsymbol z$ and $\boldsymbol U\boldsymbol z$ are excluded by additionally requiring, for $a\in\{V,U\}$ and the same index and sign choices,
\begin{equation}
	\begin{cases}
		n_z(\sigma X_a+X_V+X_r-(2-r)X_z-X_p)>1\,, &s=1,\\
		n_z(\sigma X_a-X_V+X_r-(5-r)X_z-X_p)>1\,, &s=2.
	\end{cases}
\end{equation}
    Finally, $\boldsymbol U^2$ terms must be forbidden for operators with left-handed quarks, such as $[Q_{\ell q}]_{1212}$ and $[Q_{qe}]_{1212}$. This requirement is satisfied by
    \begin{align}
    |2X_U| \neq |X_2-X_1| \,, 
    \qquad
    |2X_U| \neq |X_2-X_1-X_z| \,.
    \end{align}
    \end{enumerate}
\end{enumerate}

\subsection{EDM flavor invariants}

In \cref{tab:EDM_invs} we collect the irreducible leading electron- and neutron-EDM invariants when CP violation is sourced only by $\boldsymbol U$, for the charges in \cref{eq:chargesBench}. Additional flavor-singlet CP breaking proportional to $\epsilon_{\rm CP}$ is not included in this table and is constrained separately in the main text.

\begin{table}[h]
\input{Tab_EDM_sup}
\caption{Leading flavor invariants relevant for the electron and neutron EDMs, with the associated constraints shown in Fig.~\ref{fig:bounds}, assuming that CP violation is encoded solely in the spurion $\boldsymbol{U}$; see \cref{eq:4}.}
\label{tab:EDM_invs}
\end{table}

\clearpage
\subsection{SMEFT operator basis up to second order in spurion expansion}

\begin{table}[h]
\input{Tab_SMEFT_basis_4F_1}
\caption{Non-redundant flavor invariants associated with the $(\bar LL)(\bar LL)$ and $(\bar RR)(\bar RR)$ operator classes, constructed up to second order in spurion expansion for the $SU(2)_q \times U(1)_X$ flavor symmetry with charges given in \cref{eq:chargesBench}. The first column specifies the SMEFT class, the second lists the operators, the third indicates the spurion order, the fourth collects the corresponding flavor invariants, and the fifth (sixth) reports the number of CP-even (CP-odd) independent parameters. Entries highlighted with red boxes denote the $\cO(1)$ flavor-charged currents that arise in the $SU(2)_q \times U(1)_X$ case.}
\end{table}

\begin{table}[h]
\input{Tab_SMEFT_basis_4F_2}
\caption{Non-redundant flavor invariants associated with the $(\bar LL)(\bar RR)$ operator class, constructed up to second order in spurion expansion for the $SU(2)_q \times U(1)_X$ flavor symmetry with charges given in \cref{eq:chargesBench}. The first column specifies the SMEFT class, the second lists the operators, the third indicates the spurion order, the fourth collects the corresponding flavor invariants, and the fifth (sixth) reports the number of CP-even (CP-odd) independent parameters. Entry highlighted with red box denotes the $\cO(1)$ flavor-charged current that arises in the $SU(2)_q \times U(1)_X$ case.}
\end{table}

\begin{table}[h]
\input{Tab_SMEFT_basis_2F}
\caption{Non-redundant flavor invariants associated with the $\psi^2 H^3$, $\psi^2 XH$ and $\psi^2 H^2 D$ operator classes, constructed up to second order in spurion expansion for the $SU(2)_q \times U(1)_X$ flavor symmetry with charges given in \cref{eq:chargesBench}. The first column specifies the SMEFT class, the second lists the operators, the third indicates the spurion order, the fourth collects the corresponding flavor invariants, and the fifth (sixth) reports the number of CP-even (CP-odd) independent parameters. The entry highlighted by the blue box denotes the operator that arises in the $SU(2)_q \times U(1)_X$ case, but has no $\cO(1)$ counterpart in the $SU(2)\times U(1)^{13}$ scenario.}
\end{table}

\begin{table}[H]
\input{Tab_SMEFT_basis_4F_3}
\caption{Non-redundant flavor invariants associated with the $(\bar LR)(\bar LR)$ and $(\bar LR)(\bar RL)$ operator classes, constructed up to second order in spurion expansion for the $SU(2)_q \times U(1)_X$ flavor symmetry with charges given in \cref{eq:chargesBench}. The first column specifies the SMEFT class, the second lists the operators, the third indicates the spurion order, the fourth collects the corresponding flavor invariants, and the fifth (sixth) reports the number of CP-even (CP-odd) independent parameters. The entry highlighted by the blue box denotes the operator that arises in the $SU(2)_q \times U(1)_X$ case, but has no $\cO(1)$ counterpart in the $SU(2)\times U(1)^{13}$ scenario.}
\end{table}

\end{widetext}

\end{document}

%% file: Tab_SMEFT_count.tex
\centering
\scalebox{0.8}{
\begin{tabular}{c|cc|cc|cc|cc}
\multirow{1}{*}{\textbf{Operator class}} & \multicolumn{2}{c|}{$\boldsymbol{U(3)^5}$} 
& \multicolumn{2}{c|}{$\boldsymbol{U(2)^5}$} 
& \multicolumn{2}{c|}{$\boldsymbol{SU(2)_q \times U(1)^{13}}$} 
& \multicolumn{2}{c}{$\boldsymbol{SU(2)_q \times U(1)_X}$} 
\\
$\sscript{\textbf{CP}}$
 & $\sscript{\textbf{even}}$ & $\sscript{\textbf{odd}}$ & $\sscript{\textbf{even}}$ & $\sscript{\textbf{odd}}$ & 
 $\sscript{\textbf{~~~even}}$ & $\sscript{\textbf{~~~odd}}$
 &
 $\sscript{\textbf{~~~even}}$ & $\sscript{\textbf{~~~odd}}$
 \\ 
 \hline
 {\rule{0pt}{1.2em}\textbf{Bosonic}}
 &9&6&9&6&~~9&~~6&~~9&~~6
 \\[0.1cm]
$\boldsymbol{\psi^2 H^3}$ 
& - & - & 3 & 3&~~1&~~1& ~~1 & ~~1
\\[0.1cm]
$\boldsymbol{\psi^2 X H}$ 
& - & - & 8 & 8 &~~3&~~3& ~~3 & ~~3 
\\[0.1cm]
$\boldsymbol{\psi^2 H^2 D}$ 
& 7 & - & 15 & 1 &~~19&~~-& ~~20 & ~~1 
\\[0.1cm]
$\boldsymbol{(\bar LL)(\bar LL)}$ 
& 8 & - & 23 & - &~~31&~~-&~~32 &~~1
\\[0.1cm]
$\boldsymbol{(\bar RR)(\bar RR)}$ 
& 9 & - & 29 & - &~~60&~~-&~~61 &~~1 
\\[0.1cm]
$\boldsymbol{(\bar LL)(\bar RR)}$ 
& 8 & - & 32 & - &~~57&~~-&~~58 & ~~1
\\[0.1cm]
$\boldsymbol{(\bar LR)(\bar RL)}$ 
& - & - & 1 & 1 &~~-&~~-& ~~1 & ~~1 
\\[0.1cm]
$\boldsymbol{(\bar LR)(\bar LR)}$ 
& - & - & 4 & 4 &~~-&~~-& ~~- & ~~-
\\[0.1cm]
\hline
{\rule{0pt}{1.2em} \textbf{Total}} 
& 41 & 6 & 124 & 23 &~~180&~~10& ~~185 & ~~15
\end{tabular}
}

%% file: Tab_EDM_sup.tex
\centering
\scalebox{1.0}{
\begin{tabular}{
    c@{\hspace{0.9cm}}c@{\hspace{0.9cm}}c}
\toprule
\textbf{Operator} & 
\textbf{Order} &
\textbf{Invariants}
\\
\midrule
$[Q_{e(B,W)}]_{11}$
&$\cO(\boldsymbol U \boldsymbol V^2 \boldsymbol W \boldsymbol z^2)$
&$\boldsymbol{z}^2(\boldsymbol{U}^\dagger \boldsymbol{V})(\boldsymbol{W}^\dagger \boldsymbol{V})[Q_{e(B,W)}]_{11}$
\\[0.1cm]
&$\cO(\boldsymbol U^2 \boldsymbol V \boldsymbol W \boldsymbol z^2)$
&$\boldsymbol{z}^2(\varepsilon^{ac}\boldsymbol{U}_a^\dagger \boldsymbol{V}_c^\dagger)(\varepsilon^{bd} \boldsymbol{U}_b^\dagger \boldsymbol{W}_d^\dagger)[Q_{e(B,W)}]_{11}$
\\[0.1cm]
\midrule
$[Q_{u(B,W,G)}]_{11}$
&$\cO(\boldsymbol{U}^2 \boldsymbol{W}\boldsymbol{z})$
&$\boldsymbol{z}(\varepsilon^{ab}\boldsymbol{U}_b^\dagger)(\varepsilon^{cd}\boldsymbol{U}_c^\dagger \boldsymbol{W}_d^\dagger)[Q_{u(B,W,G)}]_{a1}$
\\[0.1cm]
$[Q_{d(B,W,G)}]_{11}$
&$\cO(\boldsymbol{U}^2 \boldsymbol{W}\boldsymbol{z})$
&$\boldsymbol{z}(\varepsilon^{ab}\boldsymbol{U}_b^\dagger)(\varepsilon^{cd}\boldsymbol{U}_c^\dagger \boldsymbol{W}_d^\dagger)[Q_{d(B,W,G)}]_{a1}$
\\[0.1cm]
\midrule
$[Q_{\ell equ}^{(1,3)}]_{1133}$
&$\cO(\boldsymbol U \boldsymbol V^2 \boldsymbol W \boldsymbol z^2)$
&$\boldsymbol{z}^2(\boldsymbol{U}^\dagger \boldsymbol{V})(\boldsymbol{W}^\dagger \boldsymbol{V})[Q_{\ell equ}^{(1,3)}]_{1133}$
\\[0.1cm]
&$\cO(\boldsymbol U^2 \boldsymbol V \boldsymbol W \boldsymbol z^2)$
&$\boldsymbol{z}^2(\varepsilon^{ac}\boldsymbol{U}_a^\dagger \boldsymbol{V}_c^\dagger)(\varepsilon^{bd} \boldsymbol{U}_b^\dagger \boldsymbol{W}_d^\dagger)[Q_{\ell equ}^{(1,3)}]_{1133}$
\\[0.1cm]
\bottomrule
\end{tabular}
}

%% file: Tab_SMEFT_basis_4F_1.tex
\scalebox{0.715}{
\begin{tabular}{
    c@{\hspace{0.3cm}}c@{\hspace{0.3cm}}c@{\hspace{0.7cm}}c@{\hspace{0.7cm}}c@{\hspace{0.5cm}}c
}
\toprule
\multirow{2}{*}{\textbf{Class}} & 
\multirow{2}{*}{\textbf{Operator}} &
\multirow{2}{*}{\textbf{Order}}&
\multirow{2}{*}{\textbf{Invariants}}&
\textbf{\# CP} &
\textbf{\# CP}
\\
&&&&\textbf{even}&\textbf{odd}
\\
\midrule
\multirow{1}{*}{$\boldsymbol{(\bar LL)(\bar LL)}$}
&\multirow{1}{*}{$Q_{qq}^{(1,3)}$  } 
&$\cO(1)$&$[Q_{qq}^{(1,3)}]_{aabb}\oplus [Q_{qq}^{(1,3)}]_{abba}\oplus [Q_{qq}^{(1,3)}]_{3aa3}\oplus[Q_{qq}^{(1,3)}]_{33aa}\oplus [Q_{qq}^{(1,3)}]_{3333}$
&5&0
\\[0.1cm]
&&$\cO(\boldsymbol{V}^2)$
&$\boldsymbol{V}^a(\boldsymbol{V}^\dag)^c[Q_{qq}^{(1,3)}]_{abbc}\oplus \boldsymbol{V}^a(\boldsymbol{V}^\dag)^b[Q_{qq}^{(1,3)}]_{abcc}\oplus (\boldsymbol{V}^\dag)^a\boldsymbol{V}^b[Q_{qq}^{(1,3)}]_{3ab3}\oplus \boldsymbol{V}^a(\boldsymbol{V}^\dag)^b[Q_{qq}^{(1,3)}]_{33ab}$
&4&0
\\[0.1cm]
&&$\cO(\boldsymbol{W}^2)$
&$\boldsymbol{W}^a(\boldsymbol{W}^\dag)^c[Q_{qq}^{(1,3)}]_{abbc}\oplus \boldsymbol{W}^a(\boldsymbol{W}^\dag)^b[Q_{qq}^{(1,3)}]_{abcc}\oplus (\boldsymbol{W}^\dag)^a\boldsymbol{W}^b[Q_{qq}^{(1,3)}]_{3ab3}\oplus \boldsymbol{W}^a(\boldsymbol{W}^\dag)^b[Q_{qq}^{(1,3)}]_{33ab}$
&4&0
\\[0.1cm]
&&$\cO(\boldsymbol{U}^2)$
&$\boldsymbol{U}^a(\boldsymbol{U}^\dag)^c[Q_{qq}^{(1,3)}]_{abbc}\oplus \boldsymbol{U}^a(\boldsymbol{U}^\dag)^b[Q_{qq}^{(1,3)}]_{abcc}\oplus (\boldsymbol{U}^\dag)^a\boldsymbol{U}^b[Q_{qq}^{(1,3)}]_{3ab3}\oplus \boldsymbol{U}^a(\boldsymbol{U}^\dag)^b[Q_{qq}^{(1,3)}]_{33ab}$
&4&0
\\[0.1cm]
\cmidrule{2-6} 
&\multirow{1}{*}{$Q_{\ell\ell}$  }
&$\cO(1)$
&$[Q_{\ell\ell}]_{1111}\oplus[Q_{\ell\ell}]_{2222}\oplus[Q_{\ell\ell}]_{3333}\oplus[Q_{\ell\ell}]_{1122}\oplus[Q_{\ell\ell}]_{1133}$
&\multirow{2}{*}{10}&\multirow{2}{*}{1}
\\[0.1cm]
&&&$\oplus[Q_{\ell\ell}]_{2233}\oplus[Q_{\ell\ell}]_{1221}\oplus[Q_{\ell\ell}]_{1331}\oplus[Q_{\ell\ell}]_{2332}
\oplus\fcolorbox{red}{white}{$[Q_{\ell\ell}]_{3121}$}$
\\[0.1cm]
\cmidrule{3-6} 
&&$\cO(\boldsymbol{V}\boldsymbol{W})$
&$(\varepsilon_{ab}\boldsymbol{V}^a\boldsymbol{W}^b)[Q_{\ell\ell}]_{1311}
\oplus
(\varepsilon_{ab}\boldsymbol{V}^a\boldsymbol{W}^b)[Q_{\ell\ell}]_{2111}
\oplus 
(\varepsilon_{ab}\boldsymbol{V}^a\boldsymbol{W}^b)[Q_{\ell\ell}]_{2213}
\oplus 
(\varepsilon_{ab}\boldsymbol{V}^a\boldsymbol{W}^b)[Q_{\ell\ell}]_{2221}
$
\\[0.1cm]
&&&$\oplus
(\varepsilon_{ab}\boldsymbol{V}^a\boldsymbol{W}^b)[Q_{\ell\ell}]_{2312}\oplus (\varepsilon_{ab}\boldsymbol{V}^a\boldsymbol{W}^b)[Q_{\ell\ell}]_{3123}
\oplus(\varepsilon_{ab}\boldsymbol{V}^a\boldsymbol{W}^b)[Q_{\ell\ell}]_{3313}
\oplus (\varepsilon_{ab}\boldsymbol{V}^a\boldsymbol{W}^b)[Q_{\ell\ell}]_{3321}
$&10&10
\\[0.1cm]
&&& $\oplus (\boldsymbol{W}^\dag\boldsymbol{V})[Q_{\ell\ell}]_{3212}\oplus (\boldsymbol{W}^\dag\boldsymbol{V})[Q_{\ell\ell}]_{3231}$
\\[0.1cm]
\cmidrule{3-6} 
&&$\cO(\boldsymbol{U}\boldsymbol{V})$
&
$(\varepsilon_{ab} \boldsymbol{U}^a\boldsymbol{V}^b)[Q_{\ell\ell}]_{1313}
\oplus
(\varepsilon_{ab} \boldsymbol{U}^a\boldsymbol{V}^b)[Q_{\ell\ell}]_{2113}
\oplus
(\varepsilon_{ab} \boldsymbol{U}^a\boldsymbol{V}^b)[Q_{\ell\ell}]_{2121}
\oplus 
(\varepsilon_{ab} \boldsymbol{U}^a\boldsymbol{V}^b)[Q_{\ell\ell}]_{2311}
$
&\multirow{2}{*}{7}&\multirow{2}{*}{7}
\\[0.1cm]
&&&
$\oplus
(\varepsilon_{ab} \boldsymbol{U}^a\boldsymbol{V}^b)[Q_{\ell\ell}]_{2322}
\oplus
(\varepsilon_{ab} \boldsymbol{U}^a\boldsymbol{V}^b)[Q_{\ell\ell}]_{3323}
\oplus 
(\boldsymbol{V}^\dag \boldsymbol{U}) [Q_{\ell\ell}]_{2323}$
\\[0.1cm]
\cmidrule{3-6} 
&&$\cO(\boldsymbol{U}\boldsymbol{W})$
&$(\boldsymbol{W}^\dag \boldsymbol{U})[Q_{\ell\ell}]_{1311}
\oplus 
(\boldsymbol{W}^\dag \boldsymbol{U})[Q_{\ell\ell}]_{2111}
\oplus 
(\boldsymbol{W}^\dag \boldsymbol{U})[Q_{\ell\ell}]_{2213}
\oplus 
(\boldsymbol{W}^\dag \boldsymbol{U})[Q_{\ell\ell}]_{2221}$
&\multirow{2}{*}{8}&\multirow{2}{*}{8}
\\[0.1cm]
&&&$
\oplus 
(\boldsymbol{W}^\dag \boldsymbol{U})[Q_{\ell\ell}]_{2312}
\oplus 
(\boldsymbol{W}^\dag \boldsymbol{U})[Q_{\ell\ell}]_{3123}
\oplus
(\boldsymbol{W}^\dag \boldsymbol{U})[Q_{\ell\ell}]_{3313}
\oplus 
(\boldsymbol{W}^\dag \boldsymbol{U})[Q_{\ell\ell}]_{3321}$
\\[0.1cm]
\cmidrule{2-6} 
&\multirow{1}{*}{$Q_{\ell q}^{(1,3)}$  }
&$\cO(1)$
&$[Q_{\ell q}^{(1,3)}]_{11aa}\oplus[Q_{\ell q}^{(1,3)}]_{22aa}\oplus[Q_{\ell q}^{(1,3)}]_{33aa}\oplus[Q_{\ell q}^{(1,3)}]_{1133}\oplus[Q_{\ell q}^{(1,3)}]_{2233}\oplus[Q_{\ell q}^{(1,3)}]_{3333}$
&6&0
\\[0.1cm]
&&$\cO(\boldsymbol{V}^2)$
&
$(\varepsilon_{bc}\boldsymbol{V}^a\boldsymbol{V}^c)[Q_{\ell q}^{(1,3)}]_{32ab}
\oplus
\boldsymbol{V}^a(\boldsymbol{V}^\dag)^b [Q_{\ell q}^{(1,3)}]_{11ab}
\oplus 
\boldsymbol{V}^a(\boldsymbol{V}^\dag)^b [Q_{\ell q}^{(1,3)}]_{22ab}
\oplus
\boldsymbol{V}^a(\boldsymbol{V}^\dag)^b [Q_{\ell q}^{(1,3)}]_{33ab}
$
&4&1
\\[0.1cm]
\cmidrule{3-6} 
&&$\cO(\boldsymbol{V}\boldsymbol{W})$
&
$(\varepsilon_{bc}\boldsymbol{V}^b\boldsymbol{W}^c)[Q_{\ell q}^{(1,3)}]_{13aa}
\oplus (\varepsilon_{bc}\boldsymbol{V}^a\boldsymbol{W}^c)[Q_{\ell q}^{(1,3)}]_{13ab}
\oplus (\varepsilon_{bc}\boldsymbol{V}^c\boldsymbol{W}^a)[Q_{\ell q}^{(1,3)}]_{13ab}
\oplus (\varepsilon_{ab}\boldsymbol{V}^a\boldsymbol{W}^b)[Q_{\ell q}^{(1,3)}]_{1333}$
&\multirow{2}{*}{8}&\multirow{2}{*}{8}
\\[0.1cm]
&&&
$\oplus (\varepsilon_{bc}\boldsymbol{V}^b\boldsymbol{W}^c)[Q_{\ell q}^{(1,3)}]_{21aa}
\oplus (\varepsilon_{bc}\boldsymbol{V}^a\boldsymbol{W}^c)[Q_{\ell q}^{(1,3)}]_{21ab}
\oplus (\varepsilon_{bc}\boldsymbol{V}^c\boldsymbol{W}^a)[Q_{\ell q}^{(1,3)}]_{21ab}
\oplus (\varepsilon_{ab}\boldsymbol{V}^a\boldsymbol{W}^b)[Q_{\ell q}^{(1,3)}]_{2133}$
\\[0.1cm]
\cmidrule{3-6} 
&&$\cO(\boldsymbol{U}\boldsymbol{V})$
&$(\varepsilon_{bc}\boldsymbol{U}^b\boldsymbol{V}^c)[Q_{\ell q}^{(1,3)}]_{23aa}\oplus (\varepsilon_{bc}\boldsymbol{U}^a\boldsymbol{V}^c)[Q_{\ell q}^{(1,3)}]_{23ab}\oplus (\varepsilon_{bc}\boldsymbol{U}^c \boldsymbol{V}^a)[Q_{\ell q}^{(1,3)}]_{23ab}\oplus (\varepsilon_{ab}\boldsymbol{U}^a\boldsymbol{V}^b)[Q_{\ell q}^{(1,3)}]_{2333}$
&4&4
\\[0.1cm]
&&$\cO(\boldsymbol{W}^2)$
&$\boldsymbol{W}^a(\boldsymbol{W}^\dag)^b[Q_{\ell q}^{(1,3)}]_{11ab}\oplus \boldsymbol{W}^a(\boldsymbol{W}^\dag)^b[Q_{\ell q}^{(1,3)}]_{22ab}\oplus \boldsymbol{W}^a(\boldsymbol{W}^\dag)^b[Q_{\ell q}^{(1,3)}]_{33ab}$
&3&0
\\[0.1cm]
\cmidrule{3-6} 
&&$\cO(\boldsymbol{U}\boldsymbol{W})$
&$
(\boldsymbol{W}^\dag \boldsymbol{U})[Q_{\ell q}^{(1,3)}]_{13aa}\oplus \boldsymbol{U}^a(\boldsymbol{W}^\dag)^b [Q_{\ell q}^{(1,3)}]_{13ab}\oplus (\boldsymbol{W}^\dag \boldsymbol{U})[Q_{\ell q}^{(1,3)}]_{1333}
$
&\multirow{2}{*}{6}&\multirow{2}{*}{6}
\\[0.1cm]
&&& $\oplus (\boldsymbol{W}^\dag \boldsymbol{U})[Q_{\ell q}^{(1,3)}]_{21aa}\oplus \boldsymbol{U}^a (\boldsymbol{W}^\dag)^b[Q_{\ell q}^{(1,3)}]_{21ab}
\oplus (\boldsymbol{W}^\dag \boldsymbol{U})[Q_{\ell q}^{(1,3)}]_{2133}$
\\[0.1cm]
\cmidrule{3-6} 
&&$\cO(\boldsymbol{U}^2)$
&$\boldsymbol{U}^a(\boldsymbol{U}^\dag)^b[Q_{\ell q}^{(1,3)}]_{11ab}\oplus \boldsymbol{U}^a(\boldsymbol{U}^\dag)^b[Q_{\ell q}^{(1,3)}]_{22ab}\oplus \boldsymbol{U}^a(\boldsymbol{U}^\dag)^b[Q_{\ell q}^{(1,3)}]_{33ab}$
&3&0
\\[0.1cm]
\midrule
\multirow{1}{*}{$\boldsymbol{(\bar RR)(\bar RR)}$}
&\multirow{1}{*}{$Q_{uu}$  }
&$\cO(1)$
&$[Q_{uu}]_{1111}\oplus[Q_{uu}]_{2222}\oplus[Q_{uu}]_{3333}\oplus[Q_{uu}]_{1122}\oplus[Q_{uu}]_{1133}$
&\multirow{2}{*}{9}&\multirow{2}{*}{0}
\\[0.1cm]
&&& $\oplus[Q_{uu}]_{2233}\oplus[Q_{uu}]_{1221}\oplus [Q_{uu}]_{1331}\oplus [Q_{uu}]_{2332}$
\\[0.1cm]
\cmidrule{3-6} 
&&$\cO(\boldsymbol{U}\boldsymbol{V})$
&$(\varepsilon_{ab}\boldsymbol{U}^a\boldsymbol{V}^b)[Q_{uu}]_{3212}$
&1&1
\\[0.1cm]
\cmidrule{2-6} 
&\multirow{1}{*}{$Q_{dd}$  }
&$\cO(1)$
&$[Q_{dd}]_{1111}\oplus[Q_{dd}]_{2222}\oplus[Q_{dd}]_{3333}\oplus[Q_{dd}]_{1122}\oplus[Q_{dd}]_{1133}$
&\multirow{2}{*}{9}&\multirow{2}{*}{0}
\\[0.1cm]
&&& $\oplus[Q_{dd}]_{2233}\oplus[Q_{dd}]_{1221}\oplus [Q_{dd}]_{1331}\oplus [Q_{dd}]_{2332}$
\\[0.1cm]
\cmidrule{3-6} 
&&$\cO(\boldsymbol{U}\boldsymbol{V})$
&$(\varepsilon_{ab}\boldsymbol{U}^a\boldsymbol{V}^b)[Q_{dd}]_{2113}
\oplus (\varepsilon_{ab}\boldsymbol{U}^a\boldsymbol{V}^b)[Q_{dd}]_{2311}
\oplus (\varepsilon_{ab}\boldsymbol{U}^a\boldsymbol{V}^b)[Q_{dd}]_{2322}
$
&\multirow{2}{*}{5}&\multirow{2}{*}{5}
\\[0.1cm]
&&& $\oplus (\varepsilon_{ab}\boldsymbol{U}^a\boldsymbol{V}^b)[Q_{dd}]_{3323}
\oplus (\boldsymbol{V}^\dag\boldsymbol{U})[Q_{dd}]_{2323}$
\\[0.1cm]
\cmidrule{2-6} 
&\multirow{1}{*}{$Q_{ud}^{(1,8)}$  }
&$\cO(1)$
&$[Q_{ud}^{(1,8)}]_{1111}\oplus [Q_{ud}^{(1,8)}]_{1122}\oplus [Q_{ud}^{(1,8)}]_{1133}\oplus[Q_{ud}^{(1,8)}]_{2211}\oplus[Q_{ud}^{(1,8)}]_{2222}$
&\multirow{2}{*}{9}&\multirow{2}{*}{0}
\\[0.1cm]
&&&$\oplus[Q_{ud}^{(1,8)}]_{2233}\oplus[Q_{ud}^{(1,8)}]_{3311}\oplus[Q_{ud}^{(1,8)}]_{3322}\oplus[Q_{ud}^{(1,8)}]_{3333}$
\\[0.1cm]
\cmidrule{3-6} 
&&$\cO(\boldsymbol{z})$&$\boldsymbol{z}[Q_{ud}^{(1,8)}]_{3113}$
&1&1
\\[0.1cm]
&&$\cO(\boldsymbol{U}\boldsymbol{V})$
&$(\varepsilon_{ab}\boldsymbol{U}^a\boldsymbol{V}^b)[Q_{ud}^{(1,8)}]_{1123}\oplus (\varepsilon_{ab}\boldsymbol{U}^a\boldsymbol{V}^b)[Q_{ud}^{(1,8)}]_{2223}\oplus (\varepsilon_{ab}\boldsymbol{U}^a\boldsymbol{V}^b)[Q_{ud}^{(1,8)}]_{3323}$
&3&3
\\[0.1cm]
\cmidrule{2-6} 
&\multirow{1}{*}{$Q_{ee}$  }
&$\cO(1)$
&$[Q_{ee}]_{1111}\oplus[Q_{ee}]_{1122}\oplus[Q_{ee}]_{1133}\oplus[Q_{ee}]_{2222}\oplus[Q_{ee}]_{2233}\oplus[Q_{ee}]_{3333}$
&6&0
\\[0.1cm]
&&$\cO(\boldsymbol{V}\boldsymbol{W})$
&$(\boldsymbol{W}^\dag \boldsymbol{V})[Q_{ee}]_{3212}$
&1&1
\\[0.1cm]
\cmidrule{2-6} 
&\multirow{1}{*}{$Q_{eu}$  }
&$\cO(1)$
&$[Q_{eu}]_{1111}\oplus [Q_{eu}]_{1122}\oplus[Q_{eu}]_{1133}\oplus[Q_{eu}]_{2211}\oplus[Q_{eu}]_{2222}$
&\multirow{2}{*}{10}&\multirow{2}{*}{1}
\\[0.1cm]
&&&$\oplus [Q_{eu}]_{2233}\oplus [Q_{eu}]_{3311}\oplus [Q_{eu}]_{3322}\oplus [Q_{eu}]_{3333}
\oplus\fcolorbox{red}{white}{$[Q_{eu}]_{3123}$}
$
\\[0.1cm]
\cmidrule{3-6} 
&&$\cO(\boldsymbol{U}\boldsymbol{V})$
&$(\varepsilon_{ab}\boldsymbol{U}^a\boldsymbol{V}^b)[Q_{eu}]_{3112}$
&1&1
\\[0.1cm]
\cmidrule{2-6} 
&\multirow{1}{*}{$Q_{ed}$}
&$\cO(1)$
&$[Q_{ed}]_{1111}\oplus [Q_{ed}]_{1122}\oplus[Q_{ed}]_{1133}\oplus[Q_{ed}]_{2211}\oplus[Q_{ed}]_{2222}$
&\multirow{2}{*}{9}&\multirow{2}{*}{0}
\\[0.1cm]
&&& $\oplus [Q_{ed}]_{2233}\oplus [Q_{ed}]_{3311}\oplus [Q_{ed}]_{3322}\oplus [Q_{ed}]_{3333}$
\\[0.1cm]
\cmidrule{3-6} 
&&$\cO(\boldsymbol{z})$&
$\boldsymbol{z}[Q_{ed}]_{3223}$
&1&1
\\[0.1cm]
&&$\cO(\boldsymbol{U}\boldsymbol{V})$
&$(\varepsilon_{ab}\boldsymbol{U}^a\boldsymbol{V}^b)[Q_{ed}]_{1123}\oplus (\varepsilon_{ab}\boldsymbol{U}^a\boldsymbol{V}^b)[Q_{ed}]_{2223}\oplus (\varepsilon_{ab}\boldsymbol{U}^a\boldsymbol{V}^b)[Q_{ed}]_{3323}\oplus (\boldsymbol{V}^\dag \boldsymbol{U})[Q_{ed}]_{1232}$
&4&4
\\[0.1cm]
&&$\cO(\boldsymbol{z}^2)$
&$\boldsymbol{z}^2 [Q_{ed}]_{2321}$
&1&1
\\[0.1cm]
\bottomrule
\end{tabular}
}

%% file: Tab_SMEFT_basis_4F_2.tex
\scalebox{0.75}{
\begin{tabular}{
    c@{\hspace{0.3cm}}c@{\hspace{0.3cm}}c@{\hspace{0.7cm}}c@{\hspace{0.7cm}}c@{\hspace{0.5cm}}c
}
\toprule
\multirow{2}{*}{\textbf{Class}} & 
\multirow{2}{*}{\textbf{Operator}} &
\multirow{2}{*}{\textbf{Order}}&
\multirow{2}{*}{\textbf{Invariants}}&
\textbf{\# CP} &
\textbf{\# CP}
\\
&&&&\textbf{even}&\textbf{odd}
\\
\midrule
$\boldsymbol{(\bar LL)(\bar RR)}$
&$Q_{qu}^{(1,8)}$
&$\cO(1)$
&$[Q_{qu}^{(1,8)}]_{aa11}\oplus [Q_{qu}^{(1,8)}]_{aa22}\oplus[Q_{qu}^{(1,8)}]_{aa33}\oplus [Q_{qu}^{(1,8)}]_{3311}\oplus [Q_{qu}^{(1,8)}]_{3322}\oplus [Q_{qu}^{(1,8)}]_{3333}$
&6&0
\\[0.1cm]
&&$\cO(\boldsymbol{W})$&$\boldsymbol{W}^a[Q_{qu}^{(1,8)}]_{a332}$
&1&1
\\[0.1cm]
&&$\cO(\boldsymbol{V}^2)$&$\boldsymbol{V}^a(\boldsymbol{V}^\dag)^b[Q_{qu}^{(1,8)}]_{ab11}\oplus \boldsymbol{V}^a(\boldsymbol{V}^\dag)^b[Q_{qu}^{(1,8)}]_{ab22}\oplus \boldsymbol{V}^a(\boldsymbol{V}^\dag)^b[Q_{qu}^{(1,8)}]_{ab33}$
&3&0
\\[0.1cm]
&&$\cO(\boldsymbol{W}^2)$&$\boldsymbol{W}^a(\boldsymbol{W}^\dag)^b[Q_{qu}^{(1,8)}]_{ab11}\oplus \boldsymbol{W}^a(\boldsymbol{W}^\dag)^b[Q_{qu}^{(1,8)}]_{ab22}\oplus \boldsymbol{W}^a(\boldsymbol{W}^\dag)^b[Q_{qu}^{(1,8)}]_{ab33}$
&3&0
\\[0.1cm]
&&$\cO(\boldsymbol{U}^2)$&$\boldsymbol{U}^a(\boldsymbol{U}^\dag)^b[Q_{qu}^{(1,8)}]_{ab11}\oplus \boldsymbol{U}^a(\boldsymbol{U}^\dag)^b[Q_{qu}^{(1,8)}]_{ab22}\oplus \boldsymbol{U}^a(\boldsymbol{U}^\dag)^b[Q_{qu}^{(1,8)}]_{ab33}$
&3&0
\\[0.1cm]
\cmidrule{2-6}
&$Q_{qd}^{(1,8)}$
&$\cO(1)$
&$[Q_{qd}^{(1,8)}]_{aa11}\oplus[Q_{qd}^{(1,8)}]_{aa22}\oplus[Q_{qd}^{(1,8)}]_{aa33}\oplus[Q_{qd}^{(1,8)}]_{3311}\oplus [Q_{qd}^{(1,8)}]_{3322}\oplus [Q_{qd}^{(1,8)}]_{3333}$
&6&0
\\[0.1cm]
&&$\cO(\boldsymbol{V}^2)$
&
$
(\varepsilon_{bc}\boldsymbol{V}^a\boldsymbol{V}^c)[Q_{qd}^{(1,8)}]_{ab32}\oplus\boldsymbol{V}^a(\boldsymbol{V}^\dag)^b[Q_{qd}^{(1,8)}]_{ab11}\oplus \boldsymbol{V}^a(\boldsymbol{V}^\dag)^b[Q_{qd}^{(1,8)}]_{ab22}\oplus \boldsymbol{V}^a(\boldsymbol{V}^\dag)^b[Q_{qd}^{(1,8)}]_{ab33}$
&4&1
\\[0.1cm]
&&$\cO(\boldsymbol{U}\boldsymbol{V})$
&$(\varepsilon_{bc}\boldsymbol{U}^b\boldsymbol{V}^c)[Q_{qd}^{(1,8)}]_{aa23}\oplus (\varepsilon_{bc}\boldsymbol{U}^a \boldsymbol{V}^c)[Q_{qd}^{(1,8)}]_{ab23}\oplus (\varepsilon_{bc}\boldsymbol{U}^c\boldsymbol{V}^a)[Q_{qd}^{(1,8)}]_{ab23}\oplus (\varepsilon_{ab}\boldsymbol{U}^a\boldsymbol{V}^b)[Q_{qd}^{(1,8)}]_{3323}$
&4&4
\\[0.1cm]
&&$\cO(\boldsymbol{V}\boldsymbol{z})$
&$(\boldsymbol{z}\boldsymbol{V}^a)[Q_{qd}^{(1,8)}]_{a331}
\oplus (\boldsymbol{z}^* \varepsilon_{ab}\boldsymbol{V}^b)[Q_{qd}^{(1,8)}]_{3a12}$
&2&2
\\[0.1cm]
&&$\cO(\boldsymbol{W}^2)$
&$\boldsymbol{W}^a(\boldsymbol{W}^\dag)^b[Q_{qd}^{(1,8)}]_{ab11}\oplus \boldsymbol{W}^a(\boldsymbol{W}^\dag)^b[Q_{qd}^{(1,8)}]_{ab22}\oplus \boldsymbol{W}^a(\boldsymbol{W}^\dag)^b[Q_{qd}^{(1,8)}]_{ab33}$
&3&0
\\[0.1cm]
&&$\cO(\boldsymbol{U}^2)$
&$\boldsymbol{U}^a(\boldsymbol{U}^\dag)^b[Q_{qd}^{(1,8)}]_{ab11}\oplus \boldsymbol{U}^a(\boldsymbol{U}^\dag)^b[Q_{qd}^{(1,8)}]_{ab22}\oplus \boldsymbol{U}^a(\boldsymbol{U}^\dag)^b[Q_{qd}^{(1,8)}]_{ab33}$
&3&0
\\[0.1cm]
\cmidrule{2-6}
&$Q_{\ell e}$
&$\cO(1)$
&$[Q_{\ell e}]_{1111}\oplus [Q_{\ell e}]_{1122}\oplus [Q_{\ell e}]_{1133}\oplus [Q_{\ell e}]_{2211}\oplus [Q_{\ell e}]_{2222}$
&\multirow{2}{*}{9}
&\multirow{2}{*}{0}
\\[0.1cm]
&&& $\oplus [Q_{\ell e}]_{2233}\oplus [Q_{\ell e}]_{3311}\oplus [Q_{\ell e}]_{3322}\oplus [Q_{\ell e}]_{3333}$
\\[0.1cm]
\cmidrule{3-6}
&&$\cO(\boldsymbol{z})$
&$\boldsymbol{z}[Q_{\ell e}]_{1221}\oplus \boldsymbol{z}[Q_{\ell e}]_{2332}\oplus \boldsymbol{z}[Q_{\ell e}]_{3121}$
&3&3
\\[0.1cm]
\cmidrule{3-6}
&&$\cO(\boldsymbol{V}\boldsymbol{W})$
&$(\varepsilon_{ab}\boldsymbol{V}^a\boldsymbol{W}^b)[Q_{\ell e}]_{1311}
\oplus (\varepsilon_{ab}\boldsymbol{V}^a\boldsymbol{W}^b)[Q_{\ell e}]_{1322}
\oplus (\varepsilon_{ab}\boldsymbol{V}^a\boldsymbol{W}^b)[Q_{\ell e}]_{1333}$
&\multirow{2}{*}{6}&\multirow{2}{*}{6}
\\[0.1cm]
&&& $\oplus (\varepsilon_{ab}\boldsymbol{V}^a\boldsymbol{W}^b)[Q_{\ell e}]_{2111}
\oplus (\varepsilon_{ab}\boldsymbol{V}^a\boldsymbol{W}^b)[Q_{\ell e}]_{2122}
\oplus (\varepsilon_{ab}\boldsymbol{V}^a\boldsymbol{W}^b)[Q_{\ell e}]_{2133}$
\\[0.1cm]
\cmidrule{3-6}
&&$\cO(\boldsymbol{U}\boldsymbol{V})$
&
$(\varepsilon_{ab}\boldsymbol{U}^a\boldsymbol{V}^b)[Q_{\ell e}]_{2311}
\oplus (\varepsilon_{ab}\boldsymbol{U}^a\boldsymbol{V}^b)[Q_{\ell e}]_{2322}
\oplus (\varepsilon_{ab}\boldsymbol{U}^a\boldsymbol{V}^b)[Q_{\ell e}]_{2333}
\oplus (\boldsymbol{V}^\dag \boldsymbol{U})[Q_{\ell e}]_{3212}$
&4&4
\\[0.1cm]
\cmidrule{3-6}
&&
$\cO(\boldsymbol{U}\boldsymbol{W})$
&
$(\varepsilon_{ab}\boldsymbol{U}^a\boldsymbol{W}^b)[Q_{\ell e}]_{1212}
\oplus (\varepsilon_{ab}\boldsymbol{U}^a\boldsymbol{W}^b)[Q_{\ell e}]_{3112}
\oplus (\boldsymbol{W}^\dag \boldsymbol{U})[Q_{\ell e}]_{1311}
\oplus (\boldsymbol{W}^\dag \boldsymbol{U})[Q_{\ell e}]_{1322}
$
&\multirow{2}{*}{8}&\multirow{2}{*}{8}
\\[0.1cm]
&&&
$
\oplus (\boldsymbol{W}^\dag \boldsymbol{U})[Q_{\ell e}]_{1333}
\oplus (\boldsymbol{W}^\dag \boldsymbol{U})[Q_{\ell e}]_{2111}
\oplus (\boldsymbol{W}^\dag \boldsymbol{U})[Q_{\ell e}]_{2122}
\oplus (\boldsymbol{W}^\dag \boldsymbol{U})[Q_{\ell e}]_{2133}$
\\[0.1cm]
\cmidrule{3-6}
&&$\cO(\boldsymbol{z}^2)$
&$\boldsymbol{z}^2[Q_{\ell e}]_{1331}\oplus \boldsymbol{z}^2[Q_{\ell e}]_{2131}$
&2&2
\\[0.1cm]
\cmidrule{2-6}
&$Q_{\ell u}$
&$\cO(1)$
&$[Q_{\ell u}]_{1111}\oplus [Q_{\ell u}]_{1122}\oplus [Q_{\ell u}]_{1133}\oplus [Q_{\ell u}]_{2211}\oplus [Q_{\ell u}]_{2222}$
&\multirow{2}{*}{9}&\multirow{2}{*}{0}
\\[0.1cm]
&&& $\oplus [Q_{\ell u}]_{2233}\oplus [Q_{\ell u}]_{3311}\oplus [Q_{\ell u}]_{3322}\oplus [Q_{\ell u}]_{3333}$
\\[0.1cm]
\cmidrule{3-6}
&&$\cO(\boldsymbol{V}\boldsymbol{W})$
&$(\varepsilon_{ab}\boldsymbol{V}^a\boldsymbol{W}^b)[Q_{\ell u}]_{1311}
\oplus (\varepsilon_{ab}\boldsymbol{V}^a\boldsymbol{W}^b)[Q_{\ell u}]_{1322}
\oplus (\varepsilon_{ab}\boldsymbol{V}^a\boldsymbol{W}^b)[Q_{\ell u}]_{1333}$
&\multirow{2}{*}{6}&\multirow{2}{*}{6}
\\[0.1cm]
&&&$\oplus(\varepsilon_{ab}\boldsymbol{V}^a\boldsymbol{W}^b)[Q_{\ell u}]_{2111}
\oplus (\varepsilon_{ab}\boldsymbol{V}^a\boldsymbol{W}^b)[Q_{\ell u}]_{2122}
\oplus (\varepsilon_{ab}\boldsymbol{V}^a\boldsymbol{W}^b)[Q_{\ell u}]_{2133}$
\\[0.1cm]
\cmidrule{3-6}
&&$\cO(\boldsymbol{U}\boldsymbol{V})$
&$(\varepsilon_{ab}\boldsymbol{U}^a\boldsymbol{V}^b)[Q_{\ell u}]_{2311}\oplus(\varepsilon_{ab}\boldsymbol{U}^a\boldsymbol{V}^b)[Q_{\ell u}]_{2322}\oplus(\varepsilon_{ab}\boldsymbol{U}^a\boldsymbol{V}^b)[Q_{\ell u}]_{2333}$
&3&3
\\[0.1cm]
\cmidrule{3-6}
&&$\cO(\boldsymbol{U}\boldsymbol{W})$
&$(\boldsymbol{W}^\dag \boldsymbol{U})[Q_{\ell u}]_{1311}\oplus(\boldsymbol{W}^\dag \boldsymbol{U})[Q_{\ell u}]_{1322}\oplus(\boldsymbol{W}^\dag \boldsymbol{U})[Q_{\ell u}]_{1333}$
&\multirow{2}{*}{6}&\multirow{2}{*}{6}
\\[0.1cm]
&&&
$\oplus (\boldsymbol{W}^\dag \boldsymbol{U})[Q_{\ell u}]_{2111}
\oplus (\boldsymbol{W}^\dag \boldsymbol{U})[Q_{\ell u}]_{2122}
\oplus (\boldsymbol{W}^\dag \boldsymbol{U})[Q_{\ell u}]_{2133}$
\\[0.1cm]
\cmidrule{3-6}
&&$\cO(\boldsymbol{z}^2)$
&$\boldsymbol{z}^2[Q_{\ell u}]_{1332}\oplus \boldsymbol{z}^2[Q_{\ell u}]_{2132}$
&2&2
\\[0.1cm]
\cmidrule{2-6}
&$Q_{\ell d}$
&$\cO(1)$
&$[Q_{\ell d}]_{1111}\oplus [Q_{\ell d}]_{1122}\oplus [Q_{\ell d}]_{1133}\oplus [Q_{\ell d}]_{2211}\oplus [Q_{\ell d}]_{2222}$
&\multirow{2}{*}{10}&\multirow{2}{*}{1}
\\[0.1cm]
&&&
$\oplus [Q_{\ell d}]_{2233}\oplus [Q_{\ell d}]_{3311}\oplus [Q_{\ell d}]_{3322}\oplus [Q_{\ell d}]_{3333}
\oplus\fcolorbox{red}{white}{$[Q_{\ell d}]_{3223}$}
$
\\[0.1cm]
\cmidrule{3-6}
&&$\cO(\boldsymbol{V}\boldsymbol{W})$
&$
(\varepsilon_{ab}\boldsymbol{V}^a\boldsymbol{W}^b)[Q_{\ell d}]_{1223}
\oplus
(\varepsilon_{ab}\boldsymbol{V}^a\boldsymbol{W}^b)[Q_{\ell d}]_{1311}
\oplus 
(\varepsilon_{ab}\boldsymbol{V}^a\boldsymbol{W}^b)[Q_{\ell d}]_{1322}
\oplus 
(\varepsilon_{ab}\boldsymbol{V}^a\boldsymbol{W}^b)[Q_{\ell d}]_{1333}
$
\\[0.1cm]
&&&
$
\oplus 
(\varepsilon_{ab}\boldsymbol{V}^a\boldsymbol{W}^b)[Q_{\ell d}]_{2111}
\oplus
(\varepsilon_{ab}\boldsymbol{V}^a\boldsymbol{W}^b)[Q_{\ell d}]_{2122}
\oplus 
(\varepsilon_{ab}\boldsymbol{V}^a\boldsymbol{W}^b)[Q_{\ell d}]_{2133}
\oplus
(\varepsilon_{ab}\boldsymbol{V}^a\boldsymbol{W}^b)[Q_{\ell d}]_{3123}
$
&10&10
\\[0.1cm]
&&&
$\oplus
(\boldsymbol{W}^\dag\boldsymbol{V})[Q_{\ell d}]_{1323}
\oplus
(\boldsymbol{W}^\dag\boldsymbol{V})[Q_{\ell d}]_{3132}$
\\[0.1cm]
\cmidrule{3-6}
&&$\cO(\boldsymbol{U}\boldsymbol{V})$
&$(\varepsilon_{ab}\boldsymbol{U}^a\boldsymbol{V}^b)[Q_{\ell d}]_{1123}\oplus(\varepsilon_{ab}\boldsymbol{U}^a\boldsymbol{V}^b)[Q_{\ell d}]_{2223}\oplus (\varepsilon_{ab}\boldsymbol{U}^a\boldsymbol{V}^b)[Q_{\ell d}]_{2311}\oplus (\varepsilon_{ab}\boldsymbol{U}^a\boldsymbol{V}^b)[Q_{\ell d}]_{2322}$
&\multirow{2}{*}{7}&\multirow{2}{*}{7}
\\[0.1cm]
&&&
$\oplus (\varepsilon_{ab}\boldsymbol{U}^a\boldsymbol{V}^b)[Q_{\ell d}]_{2333}\oplus (\varepsilon_{ab}\boldsymbol{U}^a\boldsymbol{V}^b)[Q_{\ell d}]_{3323}\oplus (\boldsymbol{V}^\dag \boldsymbol{U})[Q_{\ell d}]_{2323}$
\\[0.1cm]
\cmidrule{3-6}
&&$\cO(\boldsymbol{U}\boldsymbol{W})$
&
$
(\boldsymbol{W}^\dag \boldsymbol{U})[Q_{\ell d}]_{1223}
\oplus
(\boldsymbol{W}^\dag \boldsymbol{U})[Q_{\ell d}]_{1311}
\oplus
(\boldsymbol{W}^\dag \boldsymbol{U})[Q_{\ell d}]_{1322}
\oplus
(\boldsymbol{W}^\dag \boldsymbol{U})[Q_{\ell d}]_{1333}$
&\multirow{2}{*}{8}
&\multirow{2}{*}{8}
\\[0.1cm]
&&& $\oplus
(\boldsymbol{W}^\dag \boldsymbol{U})[Q_{\ell d}]_{2111}
\oplus
(\boldsymbol{W}^\dag \boldsymbol{U})[Q_{\ell d}]_{2122}
\oplus
(\boldsymbol{W}^\dag \boldsymbol{U})[Q_{\ell d}]_{2133}
\oplus
(\boldsymbol{W}^\dag \boldsymbol{U})[Q_{\ell d}]_{3123}$
\\[0.1cm]
\cmidrule{2-6}
&$Q_{qe}$
&$\cO(1)$
&$[Q_{qe}]_{aa11}\oplus[Q_{qe}]_{aa22}\oplus[Q_{qe}]_{aa33}\oplus[Q_{qe}]_{3311}\oplus[Q_{qe}]_{3322}\oplus [Q_{qe}]_{3333}$
&6&0
\\[0.1cm]
&&$\cO(\boldsymbol{W})$&$\boldsymbol{W}^a[Q_{qe}]_{a331}$
&1&1
\\[0.1cm]
&&$\cO(\boldsymbol{V}^2)$
&$\boldsymbol{V}^a(\boldsymbol{V}^\dag)^b[Q_{qe}]_{ab11}\oplus \boldsymbol{V}^a(\boldsymbol{V}^\dag)^b[Q_{qe}]_{ab22}\oplus \boldsymbol{V}^a(\boldsymbol{V}^\dag)^b[Q_{qe}]_{ab33}$
&3&0
\\[0.1cm]
&&$\cO(\boldsymbol{V}\boldsymbol{z})$
&$(\boldsymbol{z}\varepsilon_{ab}\boldsymbol{V}^b)[Q_{qe}]_{3a23}$
&1&1
\\[0.1cm]
&&$\cO(\boldsymbol{W}^2)$
&$\boldsymbol{W}^a(\boldsymbol{W}^\dag)^b [Q_{qe}]_{ab11}\oplus \boldsymbol{W}^a(\boldsymbol{W}^\dag)^b [Q_{qe}]_{ab22}\oplus \boldsymbol{W}^a(\boldsymbol{W}^\dag)^b [Q_{qe}]_{ab33}$
&3&0
\\[0.1cm]
&&$\cO(\boldsymbol{U}^2)$
&$(\varepsilon_{bc}\boldsymbol{U}^a \boldsymbol{U}^c)[Q_{qe}]_{ab12}\oplus \boldsymbol{U}^a(\boldsymbol{U}^\dag)^b [Q_{qe}]_{ab11}\oplus \boldsymbol{U}^a(\boldsymbol{U}^\dag)^b [Q_{qe}]_{ab22}\oplus \boldsymbol{U}^a(\boldsymbol{U}^\dag)^b [Q_{qe}]_{ab33}$
&4&1
\\[0.1cm]
\bottomrule
\end{tabular}
}

%% file: Tab_SMEFT_basis_2F.tex
\centering
\scalebox{1.0}{
\begin{tabular}{
    c@{\hspace{0.3cm}}c@{\hspace{0.3cm}}c@{\hspace{0.7cm}}c@{\hspace{0.7cm}}c@{\hspace{0.5cm}}c
}
\toprule
\multirow{2}{*}{\textbf{Class}} & 
\multirow{2}{*}{\textbf{Operator}} &
\multirow{2}{*}{\textbf{Order}}&
\multirow{2}{*}{\textbf{Invariants}}&
\textbf{\# CP} &
\textbf{\# CP}
\\
&&&&\textbf{even}&\textbf{odd}
\\
\midrule
\multirow{1}{*}{$\boldsymbol{\psi^2H^3}$}
&\multirow{1}{*}{$Q_{uH}$  }
&$\cO(1)$&$[Q_{uH}]_{33}$
&1&1
\\[0.1cm]
&&$\cO(\boldsymbol{W})$&$\boldsymbol{W}^a [Q_{uH}]_{a2}$
&1&1
\\[0.1cm]
\cmidrule{2-6}
&\multirow{1}{*}{$Q_{dH}$  }
&$\cO(\boldsymbol{z})$
&$\boldsymbol{z}[Q_{dH}]_{33}$
&1&1
\\[0.1cm]
&&$\cO(\boldsymbol{V}\boldsymbol{z})$&$(\boldsymbol{z}^*\widetilde{\boldsymbol{V}}^a)[Q_{dH}]_{a2}$
&1&1
\\[0.1cm]
&&$\cO(\boldsymbol{U}\boldsymbol{z})$&$(\boldsymbol{z}^* \boldsymbol{U}^a)[Q_{dH}]_{a3}$
&1&1
\\[0.1cm]
\cmidrule{2-6}
&\multirow{1}{*}{$Q_{eH}$  }
&$\cO(\boldsymbol{z})$&$\boldsymbol{z}[Q_{eH}]_{33}$
&1&1
\\[0.1cm]
&&$\cO(\boldsymbol{U}\boldsymbol{W})$&$(\boldsymbol{U}^\dag\widetilde{\boldsymbol{W}})[Q_{eH}]_{23}$
&1&1
\\[0.1cm]
&&$\cO(\boldsymbol{z}^2)$&$\boldsymbol{z}^2[Q_{eH}]_{22}$
&1&1
\\[0.1cm]
\midrule
\multirow{1}{*}{$\boldsymbol{\psi^2XH}$}
&\multirow{1}{*}{$Q_{u(B,W,G)}$  }
&$\cO(1)$&$[Q_{u(B,W,G)}]_{33}$
&1&1
\\[0.1cm]
&&$\cO(\boldsymbol{W})$&$\boldsymbol{W}^a [Q_{u(B,W,G)}]_{a2}$
&1&1
\\[0.1cm]
\cmidrule{2-6}
&\multirow{1}{*}{$Q_{d(B,W,G)}$  }
&$\cO(\boldsymbol{z})$
&$\boldsymbol{z}[Q_{d(B,W,G)}]_{33}$
&1&1
\\[0.1cm]
&&$\cO(\boldsymbol{V}\boldsymbol{z})$&$(\boldsymbol{z}^*\widetilde{\boldsymbol{V}}^a)[Q_{d(B,W,G)}]_{a2}$
&1&1
\\[0.1cm]
&&$\cO(\boldsymbol{U}\boldsymbol{z})$&$(\boldsymbol{z}^* \boldsymbol{U}^a)[Q_{d(B,W,G)}]_{a3}$
&1&1
\\[0.1cm]
\cmidrule{2-6}
&\multirow{1}{*}{$Q_{e(B,W)}$  }
&$\cO(\boldsymbol{z})$&$\boldsymbol{z}[Q_{e(B,W)}]_{33}$
&1&1
\\[0.1cm]
&&$\cO(\boldsymbol{U}\boldsymbol{W})$&$(\boldsymbol{U}^\dag\widetilde{\boldsymbol{W}})[Q_{e(B,W)}]_{23}$
&1&1
\\[0.1cm]
&&$\cO(\boldsymbol{z}^2)$&$\boldsymbol{z}^2[Q_{e(B,W)}]_{22}$
&1&1
\\[0.1cm]
\midrule
\multirow{1}{*}{$\boldsymbol{\psi^2H^2D}$}
&\multirow{1}{*}{$Q_{Hq}^{(1,3)}$  }
&$\cO(1)$&$[Q_{Hq}^{(1,3)}]_{aa}\oplus[Q_{Hq}^{(1,3)}]_{33}$
&2&0
\\[0.1cm]
&&$\cO(\boldsymbol{V}^2)$&$\boldsymbol{V}^a(\boldsymbol{V}^\dag)^b[Q_{Hq}^{(1,3)}]_{ab}$
&1&0
\\[0.1cm]
&&$\cO(\boldsymbol{W}^2)$&$\boldsymbol{W}^a(\boldsymbol{W}^\dag)^b[Q_{Hq}^{(1,3)}]_{ab}$
&1&0
\\[0.1cm]
&&$\cO(\boldsymbol{U}^2)$&$\boldsymbol{U}^a(\boldsymbol{U}^\dag)^b[Q_{Hq}^{(1,3)}]_{ab}$
&1&0
\\[0.1cm]
\cmidrule{2-6}
&$Q_{Hu}$  
&$\cO(1)$&$[Q_{Hu}]_{11}\oplus[Q_{Hu}]_{22}\oplus[Q_{Hu}]_{33}$
&3&0
\\[0.1cm]
\cmidrule{2-6}
&\multirow{1}{*}{$Q_{Hd}$  }
&$\cO(1)$&$[Q_{Hd}]_{11}\oplus[Q_{Hd}]_{22}\oplus[Q_{Hd}]_{33}$
&3&0
\\[0.1cm]
&&$\cO(\boldsymbol{U}\boldsymbol{V})$&$(\varepsilon_{ab}\boldsymbol{U}^a\boldsymbol{V}^b)[Q_{Hd}]_{23}$
&1&1
\\[0.1cm]
\cmidrule{2-6}
&\multirow{1}{*}{$Q_{Hud}$  }
&$\cO(1)$&$\fcolorbox{blue}{white}{$[Q_{Hud}]_{11}$}$
&1&1
\\[0.1cm]
&&$\cO(\boldsymbol{z})$&$\boldsymbol{z}[Q_{Hud}]_{33}$
&1&1
\\[0.1cm]
\cmidrule{2-6}
&\multirow{1}{*}{$Q_{H\ell}^{(1,3)}$  }
&$\cO(1)$&$[Q_{H\ell}^{(1,3)}]_{11}\oplus[Q_{H\ell}^{(1,3)}]_{22}\oplus[Q_{H\ell}^{(1,3)}]_{33}$
&3&0
\\[0.1cm]
&&$\cO(\boldsymbol{V}\boldsymbol{W})$
&$(\varepsilon_{ab}\boldsymbol{V}^a\boldsymbol{W}^b)[Q_{H\ell}^{(1,3)}]_{13}\oplus(\varepsilon_{ab}\boldsymbol{V}^a\boldsymbol{W}^b)[Q_{H\ell}^{(1,3)}]_{21}$
&2&2
\\[0.1cm]
&&$\cO(\boldsymbol{U}\boldsymbol{V})$&$(\varepsilon_{ab}\boldsymbol{U}^a\boldsymbol{V}^b)[Q_{H\ell}^{(1,3)}]_{23}$
&1&1
\\[0.1cm]
&&$\cO(\boldsymbol{U}\boldsymbol{W})$&$(\boldsymbol{W}^\dagger \boldsymbol{U})[Q_{H\ell}^{(1,3)}]_{13}\oplus (\boldsymbol{W}^\dagger \boldsymbol{U})[Q_{H\ell}^{(1,3)}]_{21}$
&2&2
\\[0.1cm]
\cmidrule{2-6}
&\multirow{1}{*}{$Q_{He}$  }
&$\cO(1)$&$[Q_{He}]_{11}\oplus[Q_{He}]_{22}\oplus[Q_{He}]_{33}$
&3&0
\\[0.1cm]
\bottomrule
\end{tabular}
}

%% file: Tab_SMEFT_basis_4F_3.tex
\scalebox{0.9}{
\begin{tabular}{
    c@{\hspace{0.3cm}}c@{\hspace{0.3cm}}c@{\hspace{0.7cm}}c@{\hspace{0.7cm}}c@{\hspace{0.5cm}}c
}
\toprule
\multirow{2}{*}{\textbf{Class}} & 
\multirow{2}{*}{\textbf{Operator}} &
\multirow{2}{*}{\textbf{Order}}&
\multirow{2}{*}{\textbf{Invariants}}&
\textbf{\# CP} &
\textbf{\# CP}
\\
&&&&\textbf{even}&\textbf{odd}
\\
\midrule
$\boldsymbol{(\bar LR)(\bar LR)}$
&$Q_{quqd}^{(1,8)}$
&$\cO(\boldsymbol{z})$&$(\boldsymbol{z}\varepsilon_{ab})[Q_{quqd}^{(1,8)}]_{a1b2}\oplus \boldsymbol{z}[Q_{quqd}^{(1,8)}]_{3333}$
&2&2
\\[0.1cm]
&&$\cO(\boldsymbol{V}^2)$&$\widetilde{\boldsymbol{V}}^a \widetilde{\boldsymbol{V}}^b [Q_{quqd}^{(1,8)}]_{a3b1}$
&1&1
\\[0.1cm]
&&$\cO(\boldsymbol{U}\boldsymbol{V})$
&$\boldsymbol{U}^a\boldsymbol{V}^b [Q_{quqd}^{(1,8)}]_{a3b1}\oplus \boldsymbol{U}^b\boldsymbol{V}^a [Q_{quqd}^{(1,8)}]_{a3b1}$
&2&2
\\[0.1cm]
&&$\cO(\boldsymbol{V}\boldsymbol{z})$
&$\boldsymbol{z}^*\widetilde{\boldsymbol{V}}^a [Q_{quqd}^{(1,8)}]_{a332}\oplus \boldsymbol{z}^*\widetilde{\boldsymbol{V}}^a [Q_{quqd}^{(1,8)}]_{33a2}$
&2&2
\\[0.1cm]
&&$\cO(\boldsymbol{U}\boldsymbol{W})$
&$\varepsilon_{ab}(\boldsymbol{U}^\dagger \widetilde{\boldsymbol{W}})[Q_{quqd}^{(1,8)}]_{a1b3}\oplus \widetilde{\boldsymbol{U}}^a \widetilde{\boldsymbol{W}}^b [Q_{quqd}^{(1,8)}]_{a1b3}\oplus \widetilde{\boldsymbol{U}}^b\widetilde{\boldsymbol{W}}^a  [Q_{quqd}^{(1,8)}]_{a1b3}$
&3&3
\\[0.1cm]
&&$\cO(\boldsymbol{W}\boldsymbol{z})$
&$\boldsymbol{z}\boldsymbol{W}^a [Q_{quqd}^{(1,8)}]_{a233}\oplus \boldsymbol{z}\boldsymbol{W}^a [Q_{quqd}^{(1,8)}]_{32a3}$
&2&2
\\[0.1cm]
&&$\cO(\boldsymbol{U}\boldsymbol{z})$
&$\boldsymbol{z}^*\boldsymbol{U}^a [Q_{quqd}^{(1,8)}]_{a333}\oplus \boldsymbol{z}^*\boldsymbol{U}^a [Q_{quqd}^{(1,8)}]_{33a3}$
&2&2
\\[0.1cm]
\cmidrule{2-6}
&$Q_{\ell equ}^{(1,3)}$
&$\cO(\boldsymbol{V})$
&$\widetilde{\boldsymbol{V}}^a [Q_{\ell equ}^{(1,3)}]_{32a3}$
&1&1
\\[0.1cm]
&&$\cO(\boldsymbol{W})$
&$\boldsymbol{W}^a [Q_{\ell equ}^{(1,3)}]_{12a3}$
&1&1
\\[0.1cm]
&&$\cO(\boldsymbol{U})$
&$\boldsymbol{U}^a [Q_{\ell equ}^{(1,3)}]_{22a3}\oplus \widetilde{\boldsymbol{U}}^a [Q_{\ell equ}^{(1,3)}]_{21a3}\oplus \widetilde{\boldsymbol{U}}^a [Q_{\ell equ}^{(1,3)}]_{23a2}$
&3&3
\\[0.1cm]
&&$\cO(\boldsymbol{z})$
&$\boldsymbol{z}[Q_{\ell equ}^{(1,3)}]_{3333}$
&1&1
\\[0.1cm]
&&$\cO(\boldsymbol{U}\boldsymbol{W})$
&$(\boldsymbol{U}^\dagger \widetilde{\boldsymbol{W}})[Q_{\ell equ}^{(1,3)}]_{2333}$
&1&1
\\[0.1cm]
&&$\cO(\boldsymbol{W}\boldsymbol{z})$
&$\boldsymbol{z}\boldsymbol{W}^a [Q_{\ell equ}^{(1,3)}]_{31a3}\oplus \boldsymbol{z}\boldsymbol{W}^a [Q_{\ell equ}^{(1,3)}]_{33a2}$
&2&2
\\[0.1cm]
&&$\cO(\boldsymbol{U}\boldsymbol{z})$
&$\boldsymbol{z}\boldsymbol{U}^a [Q_{\ell equ}^{(1,3)}]_{11a3}\oplus \boldsymbol{z}\boldsymbol{U}^a[Q_{\ell equ}^{(1,3)}]_{13a2}\oplus \boldsymbol{z}^*\boldsymbol{U}^a [Q_{\ell equ}^{(1,3)}]_{33a3}$
&3&3
\\[0.1cm]
&&$\cO(\boldsymbol{z}^2)$
&$\boldsymbol{z}^2[Q_{\ell equ}^{(1,3)}]_{2233}$
&1&1
\\[0.1cm]
\midrule
$\boldsymbol{(\bar LR)(\bar RL)}$
&$Q_{\ell edq}$
&$\cO(1)$
&$\fcolorbox{blue}{white}{$[Q_{\ell edq}]_{3333}$}$
&1&1
\\[0.1cm]
&&$\cO(\boldsymbol{V})$
&$(\varepsilon_{ab}\boldsymbol{V}^b) [Q_{\ell edq}]_{321a}\oplus
(\boldsymbol{V}^\dag)^a [Q_{\ell edq}]_{221a}$
&2&2
\\[0.1cm]
&&$\cO(\boldsymbol{z})$
&$\boldsymbol{z}[Q_{\ell edq}]_{2233}\oplus \boldsymbol{z}[Q_{\ell edq}]_{3223}$
&2&2
\\[0.1cm]
&&$\cO(\boldsymbol{V}\boldsymbol{W})$
&$ (\varepsilon_{ab}\boldsymbol{V}^a\boldsymbol{W}^b)[Q_{\ell edq}]_{1333} \oplus(\boldsymbol{V}^\dagger \boldsymbol{W})[Q_{\ell edq}]_{1323}$
&2&2
\\[0.1cm]
&&$\cO(\boldsymbol{U}\boldsymbol{V})$
&$(\varepsilon_{ab}\boldsymbol{U}^a\boldsymbol{V}^b)[Q_{\ell edq}]_{2333}\oplus (\varepsilon_{ab}\boldsymbol{U}^a\boldsymbol{V}^b)[Q_{\ell edq}]_{3323}\oplus (\boldsymbol{V}^\dagger \boldsymbol{U})[Q_{\ell edq}]_{2323}$
&3&3
\\[0.1cm]
&&$\cO(\boldsymbol{V}\boldsymbol{z})$
&$\boldsymbol{z}(\boldsymbol{V}^\dagger)^a [Q_{\ell edq}]_{111a}\oplus \boldsymbol{z}^*(\boldsymbol{V}^\dagger)^a [Q_{\ell edq}]_{331a}$
&2&2
\\[0.1cm]
&&$\cO(\boldsymbol{U}\boldsymbol{W})$
&$(\boldsymbol{W}^\dagger \boldsymbol{U})[Q_{\ell edq}]_{1333}\oplus (\boldsymbol{U}^\dagger \widetilde{\boldsymbol{W}})[Q_{\ell edq}]_{2223}$
&2&2
\\[0.1cm]
&&$\cO(\boldsymbol{W}\boldsymbol{z})$
&$(\boldsymbol{z}\varepsilon_{ab}\boldsymbol{W}^b) [Q_{\ell edq}]_{211a}
\oplus
(\boldsymbol{z}^*\varepsilon_{ab}\boldsymbol{W}^b) [Q_{\ell edq}]_{131a}$
&2&2
\\[0.1cm]
&&$\cO(\boldsymbol{U}\boldsymbol{z})$
&$(\boldsymbol{z}^*\varepsilon_{ab}\boldsymbol{U}^b) [Q_{\ell edq}]_{231a}$
&1&1
\\[0.1cm]
&&$\cO(\boldsymbol{z}^2)$
&$\boldsymbol{z}^2[Q_{\ell edq}]_{1133}$
&1&1
\\[0.1cm]
\bottomrule
\end{tabular}
}

%% file: paper.bbl
\providecommand{\href}[2]{#2}\begingroup\raggedright\begin{thebibliography}{100}

\bibitem{Feruglio:2015jfa}
F.~Feruglio, {\it {Pieces of the Flavour Puzzle}},  {\em Eur. Phys. J. C} {\bf 75} (2015), no.~8 373, [\href{http://arxiv.org/abs/1503.04071}{{\tt arXiv:1503.04071}}].

\bibitem{Altmannshofer:2024hmr}
W.~Altmannshofer and A.~Greljo, {\it {Recent Progress in Flavor Model Building}},  {\em Ann. Rev. Nucl. Part. Sci.} {\bf 75} (2025), no.~1 201--322, [\href{http://arxiv.org/abs/2412.04549}{{\tt arXiv:2412.04549}}].

\bibitem{Isidori:2025iyu}
G.~Isidori, {\it {Flavour Physics and CP Violation}},  3, 2025.
\newblock \href{http://arxiv.org/abs/2503.14042}{{\tt arXiv:2503.14042}}.

\bibitem{Nir:2020jtr}
Y.~Nir, {\it {Flavour physics and CP violation}},  {\em CERN Yellow Rep. School Proc.} {\bf 5} (2020) 79--128.

\bibitem{Altmannshofer:2025rxc}
W.~Altmannshofer and P.~Stangl, {\it {Flavour Physics Beyond the Standard Model}},  \href{http://arxiv.org/abs/2508.03950}{{\tt arXiv:2508.03950}}.

\bibitem{Buras:2020xsm}
A.~Buras, {\em {Gauge Theory of Weak Decays}}.
\newblock Cambridge University Press, 6, 2020.

\bibitem{Zupan:2019uoi}
J.~Zupan, {\it {Introduction to flavour physics}},  {\em CERN Yellow Rep. School Proc.} {\bf 6} (2019) 181--212, [\href{http://arxiv.org/abs/1903.05062}{{\tt arXiv:1903.05062}}].

\bibitem{Isidori:2010kg}
G.~Isidori, Y.~Nir, and G.~Perez, {\it {Flavor Physics Constraints for Physics Beyond the Standard Model}},  {\em Ann. Rev. Nucl. Part. Sci.} {\bf 60} (2010) 355, [\href{http://arxiv.org/abs/1002.0900}{{\tt arXiv:1002.0900}}].

\bibitem{Chivukula:1987py}
R.~S. Chivukula and H.~Georgi, {\it {Composite Technicolor Standard Model}},  {\em Phys. Lett. B} {\bf 188} (1987) 99--104.

\bibitem{DAmbrosio:2002vsn}
G.~D'Ambrosio, G.~F. Giudice, G.~Isidori, and A.~Strumia, {\it {Minimal flavor violation: An Effective field theory approach}},  {\em Nucl. Phys. B} {\bf 645} (2002) 155--187, [\href{http://arxiv.org/abs/hep-ph/0207036}{{\tt hep-ph/0207036}}].

\bibitem{Cirigliano:2005ck}
V.~Cirigliano, B.~Grinstein, G.~Isidori, and M.~B. Wise, {\it {Minimal flavor violation in the lepton sector}},  {\em Nucl. Phys. B} {\bf 728} (2005) 121--134, [\href{http://arxiv.org/abs/hep-ph/0507001}{{\tt hep-ph/0507001}}].

\bibitem{Feldmann:2008ja}
T.~Feldmann and T.~Mannel, {\it {Large Top Mass and Non-Linear Representation of Flavour Symmetry}},  {\em Phys. Rev. Lett.} {\bf 100} (2008) 171601, [\href{http://arxiv.org/abs/0801.1802}{{\tt arXiv:0801.1802}}].

\bibitem{Kagan:2009bn}
A.~L. Kagan, G.~Perez, T.~Volansky, and J.~Zupan, {\it {General Minimal Flavor Violation}},  {\em Phys. Rev. D} {\bf 80} (2009) 076002, [\href{http://arxiv.org/abs/0903.1794}{{\tt arXiv:0903.1794}}].

\bibitem{Agashe:2005hk}
K.~Agashe, M.~Papucci, G.~Perez, and D.~Pirjol, {\it {Next to minimal flavor violation}},  \href{http://arxiv.org/abs/hep-ph/0509117}{{\tt hep-ph/0509117}}.

\bibitem{Egana-Ugrinovic:2018znw}
D.~Egana-Ugrinovic, S.~Homiller, and P.~Meade, {\it {Aligned and Spontaneous Flavor Violation}},  {\em Phys. Rev. Lett.} {\bf 123} (2019), no.~3 031802, [\href{http://arxiv.org/abs/1811.00017}{{\tt arXiv:1811.00017}}].

\bibitem{Barbieri:2011ci}
R.~Barbieri, G.~Isidori, J.~Jones-Perez, P.~Lodone, and D.~M. Straub, {\it {$U(2)$ and Minimal Flavour Violation in Supersymmetry}},  {\em Eur. Phys. J. C} {\bf 71} (2011) 1725, [\href{http://arxiv.org/abs/1105.2296}{{\tt arXiv:1105.2296}}].

\bibitem{Barbieri:2012uh}
R.~Barbieri, D.~Buttazzo, F.~Sala, and D.~M. Straub, {\it {Flavour physics from an approximate $U(2)^3$ symmetry}},  {\em JHEP} {\bf 07} (2012) 181, [\href{http://arxiv.org/abs/1203.4218}{{\tt arXiv:1203.4218}}].

\bibitem{Fuentes-Martin:2019mun}
J.~Fuentes-Mart{\'\i}n, G.~Isidori, J.~Pag{\`e}s, and K.~Yamamoto, {\it {With or without U(2)? Probing non-standard flavor and helicity structures in semileptonic B decays}},  {\em Phys. Lett. B} {\bf 800} (2020) 135080, [\href{http://arxiv.org/abs/1909.02519}{{\tt arXiv:1909.02519}}].

\bibitem{Grzadkowski:2010es}
B.~Grzadkowski, M.~Iskrzynski, M.~Misiak, and J.~Rosiek, {\it {Dimension-Six Terms in the Standard Model Lagrangian}},  {\em JHEP} {\bf 10} (2010) 085, [\href{http://arxiv.org/abs/1008.4884}{{\tt arXiv:1008.4884}}].

\bibitem{Isidori:2023pyp}
G.~Isidori, F.~Wilsch, and D.~Wyler, {\it {The standard model effective field theory at work}},  {\em Rev. Mod. Phys.} {\bf 96} (2024), no.~1 015006, [\href{http://arxiv.org/abs/2303.16922}{{\tt arXiv:2303.16922}}].

\bibitem{Aebischer:2025qhh}
J.~Aebischer, A.~J. Buras, and J.~Kumar, {\it {SMEFT ATLAS: The Landscape Beyond the Standard Model}},  \href{http://arxiv.org/abs/2507.05926}{{\tt arXiv:2507.05926}}.

\bibitem{Greljo:2022cah}
A.~Greljo, A.~Palavri\'c, and A.~E. Thomsen, {\it {Adding Flavor to the SMEFT}},  {\em JHEP} {\bf 10} (2022) 010, [\href{http://arxiv.org/abs/2203.09561}{{\tt arXiv:2203.09561}}].

\bibitem{Faroughy:2020ina}
D.~A. Faroughy, G.~Isidori, F.~Wilsch, and K.~Yamamoto, {\it {Flavour symmetries in the SMEFT}},  {\em JHEP} {\bf 08} (2020) 166, [\href{http://arxiv.org/abs/2005.05366}{{\tt arXiv:2005.05366}}].

\bibitem{Allwicher:2023shc}
L.~Allwicher, C.~Cornella, B.~A. Stefanek, and G.~Isidori, {\it {New Physics in the Third Generation: A Comprehensive SMEFT Analysis and Future Prospects}},  \href{http://arxiv.org/abs/2311.00020}{{\tt arXiv:2311.00020}}.

\bibitem{Note1}
The nomenclature MFP was previously used in the narrow context of warped extra dimensions~\cite {Santiago:2008vq}.

\bibitem{LHCb:2018roe}
{\bf LHCb} Collaboration, R.~Aaij et~al., {\it {Physics case for an LHCb Upgrade II - Opportunities in flavour physics, and beyond, in the HL-LHC era}},  \href{http://arxiv.org/abs/1808.08865}{{\tt arXiv:1808.08865}}.

\bibitem{Belle-II:2018jsg}
{\bf Belle-II} Collaboration, W.~Altmannshofer et~al., {\it {The Belle II Physics Book}},  {\em PTEP} {\bf 2019} (2019), no.~12 123C01, [\href{http://arxiv.org/abs/1808.10567}{{\tt arXiv:1808.10567}}]. [Erratum: PTEP 2020, 029201 (2020)].

\bibitem{Belle-II:2022cgf}
{\bf Belle-II} Collaboration, L.~Aggarwal et~al., {\it {Snowmass White Paper: Belle II physics reach and plans for the next decade and beyond}},  \href{http://arxiv.org/abs/2207.06307}{{\tt arXiv:2207.06307}}.

\bibitem{LHCb:2021glh}
{\bf LHCb} Collaboration, {\it {Framework TDR for the LHCb Upgrade II}: {Opportunities in flavour physics, and beyond, in the HL-LHC era}}, .

\bibitem{BESIII:2020nme}
{\bf BESIII} Collaboration, M.~Ablikim et~al., {\it {Future Physics Programme of BESIII}},  {\em Chin. Phys. C} {\bf 44} (2020), no.~4 040001, [\href{http://arxiv.org/abs/1912.05983}{{\tt arXiv:1912.05983}}].

\bibitem{MEGII:2018kmf}
{\bf MEG II} Collaboration, A.~M. Baldini et~al., {\it {The design of the MEG II experiment}},  {\em Eur. Phys. J. C} {\bf 78} (2018), no.~5 380, [\href{http://arxiv.org/abs/1801.04688}{{\tt arXiv:1801.04688}}].

\bibitem{Bernstein:2019fyh}
{\bf Mu2e} Collaboration, R.~H. Bernstein, {\it {The Mu2e Experiment}},  {\em Front. in Phys.} {\bf 7} (2019) 1, [\href{http://arxiv.org/abs/1901.11099}{{\tt arXiv:1901.11099}}].

\bibitem{Hesketh:2022wgw}
{\bf Mu3e} Collaboration, G.~Hesketh, S.~Hughes, A.-K. Perrevoort, and N.~Rompotis, {\it {The Mu3e Experiment}},  in {\em {Snowmass 2021}}, 4, 2022.
\newblock \href{http://arxiv.org/abs/2204.00001}{{\tt arXiv:2204.00001}}.

\bibitem{Moritsu:2022lem}
{\bf COMET} Collaboration, M.~Moritsu, {\it {Search for Muon-to-Electron Conversion with the COMET Experiment \textdagger{}}},  {\em Universe} {\bf 8} (2022), no.~4 196, [\href{http://arxiv.org/abs/2203.06365}{{\tt arXiv:2203.06365}}].

\bibitem{NA62KLEVER:2022nea}
{\bf NA62/KLEVER, US Kaon Interest Group, KOTO, LHCb} Collaboration, {\it {Searches for new physics with high-intensity kaon beams}},  in {\em {Snowmass 2021}}, 4, 2022.
\newblock \href{http://arxiv.org/abs/2204.13394}{{\tt arXiv:2204.13394}}.

\bibitem{Aebischer:2025mwl}
J.~Aebischer et~al., {\it {Kaon physics: a cornerstone for future discoveries}},  {\em J. Phys. G} {\bf 52} (2025), no.~10 100501, [\href{http://arxiv.org/abs/2503.22256}{{\tt arXiv:2503.22256}}].

\bibitem{ACME:2018yjb}
{\bf ACME} Collaboration, V.~Andreev et~al., {\it {Improved limit on the electric dipole moment of the electron}},  {\em Nature} {\bf 562} (2018), no.~7727 355--360.

\bibitem{Wu:2019jxj}
X.~Wu, Z.~Han, J.~Chow, D.~G. Ang, C.~Meisenhelder, C.~D. Panda, E.~P. West, G.~Gabrielse, J.~M. Doyle, and D.~DeMille, {\it {The metastable Q $^3\Delta_2$ state of ThO: a new resource for the ACME electron EDM search}},  {\em New J. Phys.} {\bf 22} (2020), no.~2 023013, [\href{http://arxiv.org/abs/1911.03015}{{\tt arXiv:1911.03015}}].

\bibitem{n2EDM:2021yah}
{\bf n2EDM} Collaboration, N.~J. Ayres et~al., {\it {The design of the n2EDM experiment: nEDM Collaboration}},  {\em Eur. Phys. J. C} {\bf 81} (2021), no.~6 512, [\href{http://arxiv.org/abs/2101.08730}{{\tt arXiv:2101.08730}}].

\bibitem{Greljo:2017vvb}
A.~Greljo and D.~Marzocca, {\it {High-$p_T$ dilepton tails and flavor physics}},  {\em Eur. Phys. J. C} {\bf 77} (2017), no.~8 548, [\href{http://arxiv.org/abs/1704.09015}{{\tt arXiv:1704.09015}}].

\bibitem{Allwicher:2022gkm}
L.~Allwicher, D.~A. Faroughy, F.~Jaffredo, O.~Sumensari, and F.~Wilsch, {\it {Drell-Yan tails beyond the Standard Model}},  {\em JHEP} {\bf 03} (2023) 064, [\href{http://arxiv.org/abs/2207.10714}{{\tt arXiv:2207.10714}}].

\bibitem{deBlas:2025gyz}
J.~de~Blas et~al., {\it {Physics Briefing Book: Input for the 2026 update of the European Strategy for Particle Physics}},  \href{http://arxiv.org/abs/2511.03883}{{\tt arXiv:2511.03883}}.

\bibitem{Greljo:2024ytg}
A.~Greljo, H.~Tiblom, and A.~Valenti, {\it {New physics through flavor tagging at FCC-ee}},  {\em SciPost Phys.} {\bf 18} (2025), no.~5 152, [\href{http://arxiv.org/abs/2411.02485}{{\tt arXiv:2411.02485}}].

\bibitem{Glioti:2025zpn}
A.~Glioti, D.~Marzocca, and A.~Wulzer, {\it {Flavor physics at high-energy muon colliders}},  \href{http://arxiv.org/abs/2509.08132}{{\tt arXiv:2509.08132}}.

\bibitem{Allwicher:2025bub}
L.~Allwicher, G.~Isidori, and M.~Pesut, {\it {Flavored circular collider: cornering New Physics at FCC-ee via flavor-changing processes}},  {\em Eur. Phys. J. C} {\bf 85} (2025), no.~6 631, [\href{http://arxiv.org/abs/2503.17019}{{\tt arXiv:2503.17019}}].

\bibitem{Grinstein:2010ve}
B.~Grinstein, M.~Redi, and G.~Villadoro, {\it {Low Scale Flavor Gauge Symmetries}},  {\em JHEP} {\bf 11} (2010) 067, [\href{http://arxiv.org/abs/1009.2049}{{\tt arXiv:1009.2049}}].

\bibitem{Alonso:2011yg}
R.~Alonso, M.~B. Gavela, L.~Merlo, and S.~Rigolin, {\it {On the scalar potential of minimal flavour violation}},  {\em JHEP} {\bf 07} (2011) 012, [\href{http://arxiv.org/abs/1103.2915}{{\tt arXiv:1103.2915}}].

\bibitem{Nardi:2011st}
E.~Nardi, {\it {Naturally large Yukawa hierarchies}},  {\em Phys. Rev. D} {\bf 84} (2011) 036008, [\href{http://arxiv.org/abs/1105.1770}{{\tt arXiv:1105.1770}}].

\bibitem{Buras:2011wi}
A.~J. Buras, M.~V. Carlucci, L.~Merlo, and E.~Stamou, {\it {Phenomenology of a Gauged $SU(3)^3$ Flavour Model}},  {\em JHEP} {\bf 03} (2012) 088, [\href{http://arxiv.org/abs/1112.4477}{{\tt arXiv:1112.4477}}].

\bibitem{DAgnolo:2012ulg}
R.~T. D'Agnolo and D.~M. Straub, {\it {Gauged flavour symmetry for the light generations}},  {\em JHEP} {\bf 05} (2012) 034, [\href{http://arxiv.org/abs/1202.4759}{{\tt arXiv:1202.4759}}].

\bibitem{Alonso:2013nca}
R.~Alonso, M.~B. Gavela, G.~Isidori, and L.~Maiani, {\it {Neutrino Mixing and Masses from a Minimum Principle}},  {\em JHEP} {\bf 11} (2013) 187, [\href{http://arxiv.org/abs/1306.5927}{{\tt arXiv:1306.5927}}].

\bibitem{Banks:2025baf}
H.~Banks, G.~Crawford, M.~McCullough, and D.~Sutherland, {\it {Flavour, Accidentally}},  \href{http://arxiv.org/abs/2510.03403}{{\tt arXiv:2510.03403}}.

\bibitem{Froggatt:1978nt}
C.~D. Froggatt and H.~B. Nielsen, {\it {Hierarchy of Quark Masses, Cabibbo Angles and CP Violation}},  {\em Nucl. Phys. B} {\bf 147} (1979) 277--298.

\bibitem{Leurer:1992wg}
M.~Leurer, Y.~Nir, and N.~Seiberg, {\it {Mass matrix models}},  {\em Nucl. Phys. B} {\bf 398} (1993) 319--342, [\href{http://arxiv.org/abs/hep-ph/9212278}{{\tt hep-ph/9212278}}].

\bibitem{Leurer:1993gy}
M.~Leurer, Y.~Nir, and N.~Seiberg, {\it {Mass matrix models: The Sequel}},  {\em Nucl. Phys. B} {\bf 420} (1994) 468--504, [\href{http://arxiv.org/abs/hep-ph/9310320}{{\tt hep-ph/9310320}}].

\bibitem{Fedele:2020fvh}
M.~Fedele, A.~Mastroddi, and M.~Valli, {\it {Minimal Froggatt-Nielsen textures}},  {\em JHEP} {\bf 03} (2021) 135, [\href{http://arxiv.org/abs/2009.05587}{{\tt arXiv:2009.05587}}].

\bibitem{Note2}
The Cartan of $U(3)^5$ has rank~15, corresponding to $U(1)^{15}$.

\bibitem{Lalak:2010bk}
Z.~Lalak, S.~Pokorski, and G.~G. Ross, {\it {Beyond MFV in family symmetry theories of fermion masses}},  {\em JHEP} {\bf 08} (2010) 129, [\href{http://arxiv.org/abs/1006.2375}{{\tt arXiv:1006.2375}}].

\bibitem{Cornella:2023zme}
C.~Cornella, D.~Curtin, E.~T. Neil, and J.~O. Thompson, {\it {Mapping and probing Froggatt-Nielsen solutions to the quark flavor puzzle}},  {\em Phys. Rev. D} {\bf 111} (2025), no.~1 015042, [\href{http://arxiv.org/abs/2306.08026}{{\tt arXiv:2306.08026}}].

\bibitem{Cornella:2024jaw}
C.~Cornella, D.~Curtin, G.~Krnjaic, and M.~Mellors, {\it {Testing the Froggatt-Nielsen Mechanism with Lepton Violation}},  \href{http://arxiv.org/abs/2501.00629}{{\tt arXiv:2501.00629}}.

\bibitem{Silvestrini:2018dos}
L.~Silvestrini and M.~Valli, {\it {Model-independent Bounds on the Standard Model Effective Theory from Flavour Physics}},  {\em Phys. Lett. B} {\bf 799} (2019) 135062, [\href{http://arxiv.org/abs/1812.10913}{{\tt arXiv:1812.10913}}].

\bibitem{Blum:2009sk}
K.~Blum, Y.~Grossman, Y.~Nir, and G.~Perez, {\it {Combining K0 - anti-K0 mixing and D0 - anti-D0 mixing to constrain the flavor structure of new physics}},  {\em Phys. Rev. Lett.} {\bf 102} (2009) 211802, [\href{http://arxiv.org/abs/0903.2118}{{\tt arXiv:0903.2118}}].

\bibitem{Barbieri:1995uv}
R.~Barbieri, G.~R. Dvali, and L.~J. Hall, {\it {Predictions from a U(2) flavor symmetry in supersymmetric theories}},  {\em Phys. Lett. B} {\bf 377} (1996) 76--82, [\href{http://arxiv.org/abs/hep-ph/9512388}{{\tt hep-ph/9512388}}].

\bibitem{Pomarol:1995xc}
A.~Pomarol and D.~Tommasini, {\it {Horizontal symmetries for the supersymmetric flavor problem}},  {\em Nucl. Phys. B} {\bf 466} (1996) 3--24, [\href{http://arxiv.org/abs/hep-ph/9507462}{{\tt hep-ph/9507462}}].

\bibitem{Barbieri:1997tu}
R.~Barbieri, L.~J. Hall, and A.~Romanino, {\it {Consequences of a U(2) flavor symmetry}},  {\em Phys. Lett. B} {\bf 401} (1997) 47--53, [\href{http://arxiv.org/abs/hep-ph/9702315}{{\tt hep-ph/9702315}}].

\bibitem{Antusch:2023shi}
S.~Antusch, A.~Greljo, B.~A. Stefanek, and A.~E. Thomsen, {\it {U(2) Is Right for Leptons and Left for Quarks}},  {\em Phys. Rev. Lett.} {\bf 132} (2024), no.~15 151802, [\href{http://arxiv.org/abs/2311.09288}{{\tt arXiv:2311.09288}}].

\bibitem{Linster:2018avp}
M.~Linster and R.~Ziegler, {\it {A Realistic $U(2)$ Model of Flavor}},  {\em JHEP} {\bf 08} (2018) 058, [\href{http://arxiv.org/abs/1805.07341}{{\tt arXiv:1805.07341}}].

\bibitem{Greljo:2023bix}
A.~Greljo and A.~E. Thomsen, {\it {Rising through the ranks: flavor hierarchies from a gauged SU(2) symmetry}},  {\em Eur. Phys. J. C} {\bf 84} (2024), no.~2 213, [\href{http://arxiv.org/abs/2309.11547}{{\tt arXiv:2309.11547}}].

\bibitem{Greljo:2024zrj}
A.~Greljo, A.~E. Thomsen, and H.~Tiblom, {\it {Flavor hierarchies from SU(2) flavor and quark-lepton unification}},  {\em JHEP} {\bf 08} (2024) 143, [\href{http://arxiv.org/abs/2406.02687}{{\tt arXiv:2406.02687}}].

\bibitem{Antusch:2025xrs}
S.~Antusch, K.~Hinze, and S.~Saad, {\it {Metastable cosmic strings and gravitational waves from flavor symmetry breaking}},  {\em Phys. Rev. D} {\bf 112} (2025), no.~3 035043, [\href{http://arxiv.org/abs/2503.05868}{{\tt arXiv:2503.05868}}].

\bibitem{Darme:2023nsy}
L.~Darm{\'e}, A.~Deandrea, and F.~Mahmoudi, {\it {Gauge SU(2)$_{f}$ flavour transfers}},  {\em JHEP} {\bf 05} (2024) 313, [\href{http://arxiv.org/abs/2307.09595}{{\tt arXiv:2307.09595}}].

\bibitem{Calibbi:2025rxn}
L.~Calibbi and J.~Yi, {\it {Phenomenology of Non-Abelian Gauge and Goldstone Bosons in a U(2) Flavor Model}},  \href{http://arxiv.org/abs/2511.10468}{{\tt arXiv:2511.10468}}.

\bibitem{Greljo:2026pkz}
A.~Greljo and A.~Valenti, {\it {Hierarchies from Higher Flavor Spin}},  \href{http://arxiv.org/abs/2605.15267}{{\tt arXiv:2605.15267}}.

\bibitem{Note3}
The term in the up sector $\protect \overline {q}\protect \, \protect \boldsymbol {U}(\protect \boldsymbol {z}^*)^2\protect \, u_3$ is omitted; the CKM $V_{3i}$ elements are mainly generated in the down sector. Alternatively, one could have written two terms $\protect \overline {q}\protect \, \protect \boldsymbol {U z} \protect \,d_3$ and $\protect \overline {q}\protect \, \protect \boldsymbol {U} \protect \,u_3$, in which case the charges of $q_3, u_3$ become $\vcenter {\hbox {\scalebox {0.6}[1]{$ - $}}}X_U$ while that of $d_3$ is $\vcenter {\hbox {\scalebox {0.6}[1]{$ - $}}}(X_U + X_z)$. In this case, both terms contribute similarly to the CKM elements $V_{td}$ and $V_{ts}$.

\bibitem{Note4}
The relation between $\Lambda _\protect \text {eff}$ and new mass thresholds is, of course, UV-dependent. Regarding dimension-8 operators~\cite {Murphy:2020rsh}, under flavor anarchy with $\protect \mathcal {C}_8 \sim 1$, only a small subset can probe scales as high as $\Lambda _\protect \text {eff,8} \lesssim 100$\protect \,TeV~\cite {Liao:2024xel}. However, in our setup, spurions suppressing dimension-6 also suppress all such contributions.

\bibitem{Note5}
The group is $U(1)^{13}$ instead of $U(1)^{14}$ since the top Yukawa breaks $U(1)_{q_3} \times U(1)_{u_3} \rightarrow U(1)_{q_3 + u_3}$.

\bibitem{Kley:2021yhn}
J.~Kley, T.~Theil, E.~Venturini, and A.~Weiler, {\it {Electric dipole moments at one-loop in the dimension-6 SMEFT}},  {\em Eur. Phys. J. C} {\bf 82} (2022), no.~10 926, [\href{http://arxiv.org/abs/2109.15085}{{\tt arXiv:2109.15085}}].

\bibitem{Greljo:2022jac}
A.~Greljo, J.~Salko, A.~Smolkovi{\v{c}}, and P.~Stangl, {\it {Rare b decays meet high-mass Drell-Yan}},  {\em JHEP} {\bf 05} (2023) 087, [\href{http://arxiv.org/abs/2212.10497}{{\tt arXiv:2212.10497}}].

\bibitem{Feruglio:2015gka}
F.~Feruglio, P.~Paradisi, and A.~Pattori, {\it {Lepton Flavour Violation in Composite Higgs Models}},  {\em Eur. Phys. J. C} {\bf 75} (2015), no.~12 579, [\href{http://arxiv.org/abs/1509.03241}{{\tt arXiv:1509.03241}}].

\bibitem{Davidson:2020hkf}
S.~Davidson, {\it {Completeness and complementarity for $\mu \to e\gamma \mu \to e \bar e e$ and $\mu A \to eA$}},  {\em JHEP} {\bf 02} (2021) 172, [\href{http://arxiv.org/abs/2010.00317}{{\tt arXiv:2010.00317}}].

\bibitem{Plakias:2023esq}
I.~Plakias and O.~Sumensari, {\it {Lepton flavor violation in semileptonic observables}},  {\em Phys. Rev. D} {\bf 110} (2024), no.~3 035016, [\href{http://arxiv.org/abs/2312.14070}{{\tt arXiv:2312.14070}}].

\bibitem{Delzanno:2024ooj}
F.~Delzanno, K.~Fuyuto, S.~Gonz{\`a}lez-Sol{\'\i}s, and E.~Mereghetti, {\it {Global analysis of {\ensuremath{\mu}} {\textrightarrow} e interactions in the SMEFT}},  {\em JHEP} {\bf 07} (2025) 283, [\href{http://arxiv.org/abs/2411.13497}{{\tt arXiv:2411.13497}}].

\bibitem{Calibbi:2017uvl}
L.~Calibbi and G.~Signorelli, {\it {Charged Lepton Flavour Violation: An Experimental and Theoretical Introduction}},  {\em Riv. Nuovo Cim.} {\bf 41} (2018), no.~2 71--174, [\href{http://arxiv.org/abs/1709.00294}{{\tt arXiv:1709.00294}}].

\bibitem{Panico:2016ull}
G.~Panico and A.~Pomarol, {\it {Flavor hierarchies from dynamical scales}},  {\em JHEP} {\bf 07} (2016) 097, [\href{http://arxiv.org/abs/1603.06609}{{\tt arXiv:1603.06609}}].

\bibitem{Aebischer:2020dsw}
J.~Aebischer, C.~Bobeth, A.~J. Buras, and J.~Kumar, {\it {SMEFT ATLAS of $\Delta$F = 2 transitions}},  {\em JHEP} {\bf 12} (2020) 187, [\href{http://arxiv.org/abs/2009.07276}{{\tt arXiv:2009.07276}}].

\bibitem{Belle:2022pcr}
{\bf Belle} Collaboration, S.~Watanuki et~al., {\it {Search for the Lepton Flavor Violating Decays B+{\textrightarrow}K+{\ensuremath{\tau}}{\ensuremath{\pm}}{\ensuremath{\ell}}{\ensuremath{\mp}} ({\ensuremath{\ell}}=e, {\ensuremath{\mu}}) at Belle}},  {\em Phys. Rev. Lett.} {\bf 130} (2023), no.~26 261802, [\href{http://arxiv.org/abs/2212.04128}{{\tt arXiv:2212.04128}}].

\bibitem{Brod:2022bww}
J.~Brod, J.~M. Cornell, D.~Skodras, and E.~Stamou, {\it {Global constraints on Yukawa operators in the standard model effective theory}},  {\em JHEP} {\bf 08} (2022) 294, [\href{http://arxiv.org/abs/2203.03736}{{\tt arXiv:2203.03736}}].

\bibitem{Grunwald:2025kot}
C.~Grunwald, G.~Hiller, K.~Kr{\"o}ninger, and L.~Nollen, {\it {Beyond Universality: Probing Lepton Flavor in the SMEFT}},  \href{http://arxiv.org/abs/2511.07089}{{\tt arXiv:2511.07089}}.

\bibitem{Aebischer:2018csl}
J.~Aebischer, C.~Bobeth, A.~J. Buras, and D.~M. Straub, {\it {Anatomy of $\varepsilon '/\varepsilon $ beyond the standard model}},  {\em Eur. Phys. J. C} {\bf 79} (2019), no.~3 219, [\href{http://arxiv.org/abs/1808.00466}{{\tt arXiv:1808.00466}}].

\bibitem{Alarcon:2022ero}
R.~Alarcon et~al., {\it {Electric dipole moments and the search for new physics}},  in {\em {Snowmass 2021}}, 3, 2022.
\newblock \href{http://arxiv.org/abs/2203.08103}{{\tt arXiv:2203.08103}}.

\bibitem{Gottardo:2018ptv}
{\bf ATLAS} Collaboration, C.~A. Gottardo, {\it {Search for charged lepton-flavour violation in top-quark decays at the LHC with the ATLAS detector}},  in {\em {11th International Workshop on Top Quark Physics}}, 9, 2018.
\newblock \href{http://arxiv.org/abs/1809.09048}{{\tt arXiv:1809.09048}}.

\bibitem{Bigaran:2022giz}
I.~Bigaran, X.-G. He, M.~A. Schmidt, G.~Valencia, and R.~Volkas, {\it {Lepton-flavor-violating tau decays from triality}},  {\em Phys. Rev. D} {\bf 107} (2023), no.~5 055001, [\href{http://arxiv.org/abs/2212.09760}{{\tt arXiv:2212.09760}}].

\bibitem{FCC:2025lpp}
{\bf FCC} Collaboration, M.~Benedikt et~al., {\it {Future Circular Collider Feasibility Study Report: Volume 1, Physics, Experiments, Detectors}},  \href{http://arxiv.org/abs/2505.00272}{{\tt arXiv:2505.00272}}.

\bibitem{Greljo:2023adz}
A.~Greljo and A.~Palavri\'c, {\it {Leading directions in the SMEFT}},  {\em JHEP} {\bf 09} (2023) 009, [\href{http://arxiv.org/abs/2305.08898}{{\tt arXiv:2305.08898}}].

\bibitem{Dimopoulos:1995ju}
S.~Dimopoulos and D.~W. Sutter, {\it {The Supersymmetric flavor problem}},  {\em Nucl. Phys. B} {\bf 452} (1995) 496--512, [\href{http://arxiv.org/abs/hep-ph/9504415}{{\tt hep-ph/9504415}}].

\bibitem{Isidori:2019pae}
G.~Isidori and S.~Trifinopoulos, {\it {Exploring the flavour structure of the high-scale MSSM}},  {\em Eur. Phys. J. C} {\bf 80} (2020), no.~3 291, [\href{http://arxiv.org/abs/1912.09940}{{\tt arXiv:1912.09940}}].

\bibitem{Kaplan:1991dc}
D.~B. Kaplan, {\it {Flavor at SSC energies: A New mechanism for dynamically generated fermion masses}},  {\em Nucl. Phys. B} {\bf 365} (1991) 259--278.

\bibitem{Glioti:2024hye}
A.~Glioti, R.~Rattazzi, L.~Ricci, and L.~Vecchi, {\it {Exploring the flavor symmetry landscape}},  {\em SciPost Phys.} {\bf 18} (2025), no.~6 201, [\href{http://arxiv.org/abs/2402.09503}{{\tt arXiv:2402.09503}}].

\bibitem{Stefanek:2024kds}
B.~A. Stefanek, {\it {Non-universal probes of composite Higgs models: new bounds and prospects for FCC-ee}},  {\em JHEP} {\bf 09} (2024) 103, [\href{http://arxiv.org/abs/2407.09593}{{\tt arXiv:2407.09593}}].

\bibitem{Note6}
For a different TeV-scale flavor model with chains of VLFs, see~\cite {Arkani-Hamed:2026wwy}. Unlike in our case, the predicted 1–2 mixing is sizable and would lead to excessively large meson mixing in the presence of additional generic TeV-scale dynamics.

\bibitem{Davidson:2002qv}
S.~Davidson and A.~Ibarra, {\it {A Lower bound on the right-handed neutrino mass from leptogenesis}},  {\em Phys. Lett. B} {\bf 535} (2002) 25--32, [\href{http://arxiv.org/abs/hep-ph/0202239}{{\tt hep-ph/0202239}}].

\bibitem{Greljo:2025suh}
A.~Greljo, X.~Ponce~D{\'\i}az, and A.~E. Thomsen, {\it {Insights on the cosmic origin of matter from proton stability}},  {\em JCAP} {\bf 11} (2025) 043, [\href{http://arxiv.org/abs/2505.18259}{{\tt arXiv:2505.18259}}].

\bibitem{Santiago:2008vq}
J.~Santiago, {\it {Minimal Flavor Protection: A New Flavor Paradigm in Warped Models}},  {\em JHEP} {\bf 12} (2008) 046, [\href{http://arxiv.org/abs/0806.1230}{{\tt arXiv:0806.1230}}].

\bibitem{Murphy:2020rsh}
C.~W. Murphy, {\it {Dimension-8 operators in the Standard Model Effective Field Theory}},  {\em JHEP} {\bf 10} (2020) 174, [\href{http://arxiv.org/abs/2005.00059}{{\tt arXiv:2005.00059}}].

\bibitem{Liao:2024xel}
Y.~Liao, X.-D. Ma, and H.-L. Wang, {\it {Probing dimension-8 SMEFT operators through neutral meson mixing}},  {\em JHEP} {\bf 03} (2025) 133, [\href{http://arxiv.org/abs/2409.10305}{{\tt arXiv:2409.10305}}].

\bibitem{Arkani-Hamed:2026wwy}
N.~Arkani-Hamed, C.~Figueiredo, L.~J. Hall, and C.~A. Manzari, {\it {Generating the fermion mass hierarchy at the TeV scale}},  \href{http://arxiv.org/abs/2602.17754}{{\tt arXiv:2602.17754}}.

\bibitem{Barbieri:1998em}
R.~Barbieri, L.~Giusti, L.~J. Hall, and A.~Romanino, {\it {Fermion masses and symmetry breaking of a U(2) flavor symmetry}},  {\em Nucl. Phys. B} {\bf 550} (1999) 32--40, [\href{http://arxiv.org/abs/hep-ph/9812239}{{\tt hep-ph/9812239}}].

\bibitem{Barbieri:1999pe}
R.~Barbieri, P.~Creminelli, and A.~Romanino, {\it {Neutrino mixings from a U(2) flavor symmetry}},  {\em Nucl. Phys. B} {\bf 559} (1999) 17--26, [\href{http://arxiv.org/abs/hep-ph/9903460}{{\tt hep-ph/9903460}}].

\bibitem{Wolfenstein:1983yz}
L.~Wolfenstein, {\it {Parametrization of the Kobayashi-Maskawa Matrix}},  {\em Phys. Rev. Lett.} {\bf 51} (1983) 1945.

\bibitem{ParticleDataGroup:2024cfk}
{\bf Particle Data Group} Collaboration, S.~Navas et~al., {\it {Review of particle physics}},  {\em Phys. Rev. D} {\bf 110} (2024), no.~3 030001.

\end{thebibliography}\endgroup
